\documentclass[english,aps, preprint]{revtex4}
\usepackage[T1]{fontenc}
\usepackage[latin1]{inputenc}
\usepackage{amsmath}
\usepackage{graphicx}
\usepackage{setspace}
\usepackage{amssymb}

\makeatletter
\usepackage{babel}
\makeatother
\begin{document}

\title{Statistical Mechanics of Semiflexible Polymer Chains from a new Generating
Function }

\author{Marcelo Marucho$^{a)}$, Leonardo Loureiro$^{b)}$ and Gustavo A.
Carri$^{c)}$}

\affiliation{The Maurice Morton Institute of Polymer Science, The University of
Akron, Akron, OH 44325-3909, USA.}

\begin{abstract}
In this paper, we present a new approach to the Kratky-Porod Model
(KP) of semiflexible polymers. Our solution to the model is based
on the definition of a generating function which we use to study the
statistical mechanics of semiflexible polymer chains. Specifically,
we derive mathematical expressions for the characteristic function,
the polymer propagator and the mean square end-to-end distance from
the generating function. These expressions are valid for polymers
with any number of segments and degree of rigidity. Furthermore, they
capture the limits of fully flexible and stiff polymers exactly. In
between, a smooth and approximate crossover behavior is predicted.
The most important contribution of this paper is the expression of
the polymer propagator which is written in a very simple and insightful
way. It is given in terms of the exact polymer propagator of the Random
Flight Model multiplied by an exponential that takes into account
the stiffness of the polymer backbone. 

\vspace{6.2cm}

---------------------

\begin{singlespace}
\noindent {\footnotesize Electronic address:$^{a)}$marucho@polymer.uakron.edu}{\footnotesize \par}

\noindent {\footnotesize Electronic address:$^{b)}$lla2@uakron.edu}{\footnotesize \par}

\noindent {\footnotesize Electronic address:$^{c)}$carri@polymer.uakron.edu}{\footnotesize \par}

\noindent {\footnotesize To whom any correspondence should be addressed}{\footnotesize \par}\end{singlespace}

\end{abstract}
\maketitle
\clearpage

\section{Introduction}

Since the ground-breaking ideas by Kratky and Porod in 1949,$^{1}$
theoretical studies of semiflexible polymers based on the KP model
have been abundant. In this model, the polymer chain displays resistance
to bending deformations. This resistance is modeled using a free energy
that penalizes bending the polymer backbone. The free energy depends
on parameters (elastic constants) that are a consequence of many short-range
monomer-monomer interactions. Explicitly, for the continuous version
of the KP model$^{2}$ (called the Wormlike Chain Model, WCM) the
free energy is

\begin{equation}
H=\frac{\epsilon}{2}\int_{0}^{\mathrm{L}}\mathrm{d}s\left(\frac{\partial\mathbf{R}}{\partial s}\right)^{2},\label{eq:potential_energy}\end{equation}
where $\mathbf{R}\left(\mathrm{s}\right)$ is the vectorial field
that represents the polymer chain, $s$ is the arc of length parameter,
$L$ is the contour length of the polymer and $\epsilon$ is the bending
modulus. In addition, the local inextensibility constraint $\left|\mathrm{d}\mathbf{R}\left(\mathrm{s}\right)/\mathrm{ds}\right|=1$
must be satisfied. 

As a consequence of the bending rigidity, a semiflexible chain is
characterized by a persistence length (proportional to the bending
modulus) such that, if the length scale is shorter than the persistence
length, then the chain behaves like a rod while, if the length scale
is larger than the persistence length, then the chain is governed
by the configurational entropy that favors the random-walk conformations.

The local inextensibility constraint has not allowed researchers to
find an exact solution to the model. Indeed, the constraint $\left|\mathrm{d}\mathbf{R}\left(\mathrm{s}\right)/\mathrm{ds}\right|=1$
is written using a Dirac delta distribution in infinite dimensions.
Depending on how the constraint is written, $\left|\mathrm{d}\mathbf{R}\left(\mathrm{s}\right)/\mathrm{ds}\right|=1$
or $\left(\mathrm{d}\mathbf{R}\left(\mathrm{s}\right)/\mathrm{ds}\right)^{2}=1$,
we get an Edwards Hamiltonian that is non-analytic or non-linear,
respectively. Consequently, there is no exact solution of this model
at present. However, a few properties like the first few moments of
the distribution of the end-to-end distance$^{2,3}$ are known exactly.

Indeed, many researchers have addressed the Kratky-Porod and other
models of semiflexible polymers with the purpose of understanding
the statistical behavior of this kind of polymers. Among the many
theoretical treatments of semiflexible polymers, let us start by mentioning
two very important contributions done in the 1950s: the work by Daniels$^{4}$
who developed expansions of the polymer propagator for a semiflexible
polymer in inverse powers of the number of segments and the classic
paper by Benoit and Doty$^{5}$ who obtained the exact expression
for the average end-to-end distance squared and radius of gyration.
During the following two decades, the field of statistical mechanics
of semiflexible polymers saw a substantial growth thanks to the seminal
contributions of many researchers. For example, Fixman and Kovacs
developed a modified Gaussian model for stiff polymer chains under
an external field (external force).$^{6}$ In this approach, they
computed an approximate distribution for the bond vectors from which
they were able to compute the partition function and average end-to-end
vector. An alternative approach was proposed by Harris and Hearst$^{7}$
who developed a distribution for the continuous model from which they
were able to compute the two-point correlation function and, consequently,
the mean-square end-to-end distance and radius of gyration. A reformulation
of the KP model using field-theoretic methods was developed by Sait$\hat{\mathrm{o}}$
and coworkers$^{2}$ who computed exactly different averages of the
end-to-end distance and tangent-tangent correlation function. In addition,
Gobush and coworkers$^{8}$ developed an asymptotic expression for
the polymer propagator in inverse powers of the number of segments
and, Yamakawa and Stockmayer$^{9}$ addressed the first order corrections
to the end-to-end distance squared and second virial coefficient due
to excluded volume interactions. Semiflexible polymers were also extensively
studied by Freed$^{10}$ who employed field-theoretic methods to study
their statistical behavior. A few years later, he developed a modified
Gaussian distribution approximation to the continuous version of the
KP model$^{11}$ which has been re-derived using different mathematical
methods by Lagowski and coworkers,$^{12}$ and Winkler and coworkers.$^{13}$
Similar results were obtained by Zhao and coworkers$^{14}$ who also
studied the effect of an external field. Field-theoretic methods have
also been used by Bhattacharjee and Muthukumar$^{15}$ who employed
the Edwards-Singh self-consistent approach to obtain an effective
Gaussian Hamiltonian from which they computed the average end-to-end
distance squared. 

In recent years, the advent of new experimental methods that have
allowed researchers to manipulate single molecules has generated new
momentum in the area of statistical mechanics of semiflexible polymers.$^{16}$
Indeed, new analytical approaches to semiflexible polymers have been
developed by Marko and Siggia, and Kroy and Frey$^{17}$ who derived
the force versus elongation behavior predicted by the WCM, Hansen
and Podgornik$^{18}$ who developed a non-perturbative 1/d expansion
(d being the dimension of the embedding space), Whilhem and Frey$^{19}$
who computed the polymer propagator for polymers with large bending
rigidities and Winkler$^{20}$ who computed the same quantity for
any value of the stiffness of the polymer backbone using the Maximum
Entropy Principle. Finally, it is worth mentioning one alternative
approach to semiflexible polymers (Dirac chains) developed by Kholodenko.$^{21}$

The aforementioned extensive list of approximate treatments of the
KP model, though not exhaustive, does show two major trends. On the
one hand, many studies computed different statistical properties as
perturbative corrections to the rod-like and flexible limits. One
example of this statement is the seminal paper by Gobush \textit{etal.}
where the polymer propagator was computed as a perturbation expansion
with respect to the Gaussian chain model (see Eqs. (22) and (23) in
Ref. 8). On the other hand, many researchers have targeted a few physical
properties of the KP model and optimized the approximations to capture
these properties exactly or very accurately. Two examples of this
statement are: the paper by Bawendi and Freed,$^{11}$ who optimized
their treatment of the KP model (continuous version) to reproduce
the bond-bond projection expectation value exactly and obtained a
modified Gaussian distribution for the end-to-end distance (see Eq.
(13) in Ref. 11), and the paper by Winkler \textit{etal.}$^{13}$
who used a very fundamental concept, the Maximum Entropy Principle,
to compute different ensemble averages and correlation functions (the
constraints used to optimize the use of the Maximum Entropy Principle
are given by Eqs. (3.25) to (3.28) in Ref. 13). These observations
lead naturally to the following two questions . Firstly, we should
ask {}``why do we need so many different approximations to solve
the same model?'' and, secondly, {}``can we find one approximation
capable of predicting all the statistical properties of the model
in such a way that corrections to these results can be computed in
a systematic and perturbative fashion ?''. Observe that the answer
to the latter is affirmative for the problem of a single flexible
polymer with excluded volume interaction.$^{22}$

In this paper we address the second question. In other words, we address
the KP model with the purpose of finding one {}``ground'' state
capable of reproducing all the statistical properties of the model
approximately. Furthermore, we require that our {}``ground'' state
captures all the statistical properties of the rod-like and flexible
limits exactly and respects the local, not global, inextensibility
constraint. In the crossover region, the solution is not exact but,
we require that it displays the correct physical features as described
by other treatments of the model and, moreover, it should be amenable
to systematic and controlled corrections calculated in a perturbative
manner around the {}``ground'' state. In this first paper, we set
the Fixman parameter to zero and leave the excluded volume problem
for a future publication.

The most important prediction of this calculation is the polymer propagator
which turns out to be proportional to the polymer propagator of the
Random Flight Model$^{23}$ multiplied by an exponential arising from
the stiffness of the polymer backbone described by the parameter $\alpha$
(see text for details),

\[
P\left(R\right)\sim P_{RFM}\left(R\right)\exp\left[R^{2}/2\alpha n^{2}\right],\]
where $n$ is the number of segments and $R$ is the end-to-end distance.
The mathematical simplicity of this expression makes it a very good
candidate for the development of a perturbative treatment of the KP
model.

This paper is organized as follows. In section II, we propose a generating
function and obtain an approximate analytical expression for it. This
section also contains the only approximation of our calculation. Afterward,
we use the generating function to calculate the characteristic function,
mean squared end-to-end distance and polymer propagator of the model.
In Section III we discuss the results of our calculations which are
valid for any value of the stiffness of the polymer backbone and length
of the polymer chain. For the purpose of making our presentation more
balanced and objective, we compare our results with those obtained
by other researchers. Section IV contains the conclusions of our work.
The details of some mathematical calculations are presented in the
appendix.

\section{Theory}

\subsection{Review of the Kratky-Porod Model and Evaluation of the new Generating
Function }

Let us consider a polymer chain as a set of $n$ bond vectors $\left(\mathbf{u}_{0},\mathbf{u}_{1},...,\mathbf{u}_{n-1}\right)$
connected in a sequential manner. Furthermore, let us assume that
the length of each bond vector is $l_{k}$ (=Kuhn length) and that
pairs of consecutive bond vectors try to be parallel to each other.
This orientational interaction is modeled with a Boltzmann weight
given by the following expression$^{6,15}$

\begin{equation}
{\displaystyle \exp\left(-\frac{\epsilon}{2l_{k}^{2}k_{B}T}\sum_{k=0}^{n-2}\left(\mathbf{u}_{k+1}-\mathbf{u}_{k}\right)^{2}\right)},\label{eq:potential_energy}\end{equation}
 where $\epsilon$ and $k_{B}T$ are the bending modulus and thermal
energy, respectively. In addition, we take into account the local
inextensibility constraint with the following term$^{2}$

\begin{equation}
\prod_{i=0}^{n-1}\delta\left\{ \frac{\left(\mathbf{u}_{i}\right)^{2}}{l_{k}^{2}}-1\right\} .\label{eq:inexten}\end{equation}

Equations (\ref{eq:potential_energy}) and (\ref{eq:inexten}) define
the KP model completely and all the statistical properties of the
model such as the characteristic function, single chain structure
factor and other correlation functions, probability distributions
like the polymer propagator and their moments can be calculated. The
evaluation of these statistical properties can be easily done using
the following generating function

\begin{equation}
C\left(\left\{ \Psi_{kjp}\right\} \right)\equiv\left\langle \exp\left(\sum_{k=0}^{n-1}\mathbf{u}_{k}\cdot\Psi_{kjp}\right)\right\rangle ,\label{eq:gen}\end{equation}
where $\Psi_{kjp}$ is

\begin{equation}
\Psi_{kjp}\equiv\mathbf{J}_{k}+i\mathbf{q}\cdot\left[\Theta(k-j)-\Theta(k-p)\right].\label{eq:generating}\end{equation}
This definition shows that $\Psi_{kjp}$ is an auxiliary tensor. It
represents a standard external source consisting of two terms: the
first one, $\left(\mathbf{J}_{k}\right)$, represents a dipolar coupling
with the $k$-th bond vector and, the second one, $\left(\mathbf{q}\right)$,
is also a dipolar coupling that is non-zero only when the $k$-th
bond vector is between the $j$-th and $p$-th ones. The $\Theta(z)$
function is the Heaviside step function.$^{24}$ 

If the generating function given by Eq.(\ref{eq:gen}) is known, then
all the aforementioned statistical properties can be obtained from
it by differentiation with respect to $\mathbf{J}_{k}$ and/or by
assigning specific values to $\mathbf{J}_{k}$ . For example, the
characteristic function is $C\left(\left\{ i\mathbf{q}\cdot\left[\Theta(k-j)-\Theta(k-p)\right]\right\} \right)$
from which the polymer propagator is computed as the inverse Fourier
transform of $C\left(\left\{ i\mathbf{q}\right\} \right)$, the tangent-tangent
correlation function, $\left\langle \mathbf{u}_{j}\cdot\mathbf{u}_{k}\right\rangle $,
is computed as the second derivative with respect to $\mathbf{J}_{j}$
and $\mathbf{J}_{k}$ which has to be evaluated at $\mathbf{J}_{0}=\mathbf{J}_{1}=....\mathbf{J}_{n-1}=0$.
Other statistical quantities can also be computed easily. Consequently,
our first step is to evaluate $C\left(\left\{ \Psi_{kjp}\right\} \right)$
whose explicit form is 

\begin{equation}
C\left(\left\{ \Psi_{kjp}\right\} \right)=\frac{1}{z}{\displaystyle \int}\left[\prod_{i=0}^{n-1}d\mathbf{u}_{i}\delta\left\{ \frac{\left(\mathbf{u}_{i}\right)^{2}}{l_{k}^{2}}-1\right\} \right]\exp\left\{ -\frac{\epsilon}{2l_{k}^{2}k_{B}T}\sum_{k=0}^{n-2}\left(\mathbf{u}_{k+1}-\mathbf{u}_{k}\right)^{2}+\sum_{k=0}^{n-1}\mathbf{u}_{k}\cdot\Psi_{kjp}\right\} ,\label{eq:form_of_C}\end{equation}
where $z$ is defined such that $C\left(\left\{ \mathbf{0}_{kjp}\right\} \right)=1$.
After writing the delta distributions in Eq.(\ref{eq:form_of_C})
in terms of their Fourier representations,$^{25}$ a Hubbard-Stratanovich
transformation$^{26}$ is needed in order to avoid dealing with the
quadratic term in $\mathbf{u}_{k}$ arising from the exponential representation
of the delta distributions. As a result, we get 

\begin{equation}
\begin{array}{c}
{\displaystyle C\left(\left\{ \Psi_{kjp}\right\} \right)=\frac{1}{z}{\displaystyle \int}\left[\prod_{i=0}^{n-1}d\mathbf{u}_{i}d\lambda_{i}d\Phi_{i}\right]\exp\left(-\frac{\epsilon}{2l_{k}^{2}k_{B}T}\sum_{k=0}^{n-2}\left(\mathbf{u}_{k+1}-\mathbf{u}_{k}\right)^{2}-\frac{3}{2}\sum_{k=0}^{n-1}\left[\ln\lambda_{k}+i\frac{2}{3}\lambda_{k}\right]\right.}\\
{\displaystyle \left.+\sum_{k=0}^{n-1}\mathbf{u}_{k}\cdot\left(\Psi_{kjp}-i\frac{\Phi_{k}}{l_{k}}\right)-\frac{i}{4}\sum_{k=0}^{n-1}\frac{\Phi_{k}^{2}}{\lambda_{k}}\right)},\end{array}\label{eq:stratanovich}\end{equation}
where $\lambda_{k}$ and $\Phi_{k}$ are dummy variables arising from
the Fourier representation of the delta and the Hubbard-Stratanovich
transformation, respectively. Observe that after redefining $\Psi_{kjp}l_{k}\rightarrow\Psi_{kjp}$
and $\mathbf{u}_{k}/l_{k}\rightarrow\mathbf{u}_{k}$, the dependence
of Eq.(\ref{eq:stratanovich}) on the Kuhn length goes away. 

We now proceed to compute the integrals over $\lambda_{k}$ and $\mathbf{u}_{k}$.
The $\lambda_{k}$-integrals are straightforward, the result is 

\begin{equation}
\int\left[\prod_{i=0}^{n-1}d\lambda_{i}\right]\exp\left(-\frac{i}{4}\sum_{k=0}^{n-1}\frac{\Phi_{k}^{2}}{\lambda_{k}}-\frac{3}{2}\sum_{k=0}^{n-1}\left[\ln\lambda_{k}+i\frac{2}{3}\lambda_{k}\right]\right)\sim\prod_{i=0}^{n-1}\frac{\sin[(\Phi_{i}\cdot\Phi_{i})^{\frac{1}{2}}]}{(\Phi_{i}\cdot\Phi_{i})^{\frac{1}{2}}},\label{eq:lambda_integral}\end{equation}
 where a constant prefactor independent of $\Phi_{i}$ has been neglected
because of the definition of $z$ in Eq.(\ref{eq:form_of_C}).

The $\mathbf{u}_{k}$-integrals are exactly doable using saddle point
which is given by the following expression

\begin{equation}
\begin{array}{c}
{\displaystyle \mathbf{u}_{k}=\mathbf{u}_{0}-\frac{k_{B}T}{\epsilon}\sum_{m=0}^{k-1}(\Psi_{mjp}-i\Phi_{m})\left(k-m\right),}\end{array}\label{eq:saddle_point}\end{equation}
where the following constrain must be fulfilled

\begin{equation}
\sum_{m=0}^{n-1}(\Phi_{m}+i\Psi_{mjp})=0.\label{eq:constrainsource}\end{equation}

Replacing Eqs.(\ref{eq:lambda_integral}) and (\ref{eq:saddle_point})
into Eq.(\ref{eq:stratanovich}), and expressing the constraint as
a delta distribution in the exponential representation, the generating
function takes the form 

\begin{equation}
\begin{array}{c}
{\displaystyle C\left(\left\{ \Psi_{kjp}\right\} \right)=\frac{1}{z}\int d\mathbf{u}\left\{ \int\left[\prod_{j=0}^{n-1}d\Phi_{j}\right]\exp\left[\sum_{j=0}^{n-1}\ln\left(\frac{\sin\left(\left|\Phi_{j}\right|\right)}{\left(\left|\Phi_{j}\right|\right)}\right)\right.\right.}\\
{\displaystyle \left.\left.-\frac{\kappa n}{2}\sum_{k=0}^{n-1}\sum_{s=0}^{n-1}(\Phi_{k}+i\Psi_{kjp})K_{k,\, s}(\Phi_{s}+i\Psi_{sjp})-i\sum_{s=0}^{n-1}\mathbf{u}\cdot(\Phi_{s}+i\Psi_{sjp})\right]\right\} ,}\end{array}\label{eq:Final_C}\end{equation}
where the kernel $K_{k,\, s}$ and the parameter $\kappa$ are defined
as

\begin{equation}
K_{k,\, s}\equiv\left[1-\frac{k}{n}-\frac{\Theta(s-k)}{n}(s-k)\right]\quad\textrm{and}\quad\kappa\equiv\frac{k_{B}T}{\epsilon}.\label{eq:Kernel}\end{equation}
We refer the reader to Appendix A for some mathematical details of
this part of the calculation.

The last step involves the evaluation of the $\Phi_{j}$-integrals.
Observe that this part of the calculation is not exactly doable thus,
we approximate it as follows. First, we assume that, for a fixed value
of $s$ or $k$, the element of $K_{k,\, s}$ that contributes the
most to the integral is the diagonal one. Thus, we approximate $K_{k,\, s}$
by its diagonal elements. The contributions arising from non-diagonal
elements will be treated in a future publication. We also perform
a pre-averaging approximation. Consequently, the kernel $K_{k,\, s}$
becomes

\begin{equation}
K_{k,\, s}\simeq\left\langle 1-\frac{k}{n}\right\rangle \delta_{ks}=\frac{1}{2}\left(1+\frac{1}{n}\right)\delta_{ks},\label{eq:diagonalker}\end{equation}
This is the only approximation of our work. In the next section, we
will show that this approximation captures the flexible (Random Flight
Model) and rod-like limits exactly when $\kappa\rightarrow\infty$
and $\kappa\rightarrow0$, respectively. This approximation also provides
a very good crossover behavior for intermediate values of the parameter
$\kappa$.

Equation (\ref{eq:diagonalker}) transforms the integrand of the $\mathbf{u}$-integral
into 

\begin{equation}
I_{\Phi}=\prod_{s=1}^{n-1}\int d\Phi_{s}\frac{\sin(\left|\Phi_{s}\right|)}{\left|\Phi_{s}\right|}\exp\left[-i\mathbf{u}\cdot(\Phi_{s}+i\Psi_{sjp})-\frac{\alpha n}{2}(\Phi_{s}+i\Psi_{sjp})^{2}\right],\label{eq:Int_phi}\end{equation}
where, for the purpose of simplicity, we have introduced a new parameter
$\alpha\equiv\frac{1}{2}\left(1+\frac{1}{n}\right)\kappa$. 

Finally, replacing Eq.(\ref{eq:Int_phi}) in Eq.(\ref{eq:Final_C})
we obtain the most general expression of the generating function $C\left(\left\{ \Psi_{kjp}\right\} \right)$.
In the next section we evaluate some physical properties of the KP
model using this generating function.

\subsection{Characteristic Function and the Mean Square End-to-End Distance}

Let us start by computing the characteristic function. For this purpose,
we make $\Psi=i\mathbf{q}$ in Eq.(\ref{eq:Final_C}) where $\mathbf{q}$
is the scattering wave vector. Then, the characteristic function and
the mean square end-to-end distance can be calculated. The $\Phi$-integration
takes the form

\begin{equation}
\begin{array}{c}
{\displaystyle I_{\Phi}=\prod_{j=0}^{n-1}\int_{0}^{\infty}d\Phi\int_{-1}^{1}d\left(cos\left(\theta\right)\right)\int_{0}^{2\pi}d\varphi\sin(\Phi)\Phi\exp\left[-\frac{\alpha n}{2}(\Phi^{2}+q^{2})+\right.}\\
{\displaystyle \left.+i\mathbf{q}\cdot\mathbf{u}+\alpha n\mathbf{q}\cdot\Phi-i\mathbf{u}\cdot\Phi\right]},\end{array}\label{eq:int_phi_q}\end{equation}
 whose calculation is straightforward. Choosing $\theta$ to be the
angle between the vectors $\mathbf{u}$ and $\Phi$, and $\theta_{q}$
the angle between $\mathbf{q}$ and $\Phi$, then the solution to
the $\varphi$-integral is $2\pi I_{0}\left(\alpha n\Phi q\left|\sin\left(\theta\right)\sin\left(\theta_{q}\right)\right|\right)$
where $I_{0}\left(x\right)$ is the Bessel function of second class.$^{27}$
Hence, the angular integral, $I_{\Omega}$, reads

\begin{equation}
\begin{array}{c}
{\displaystyle I_{\Omega}=2\pi\int_{-1}^{1}d\left(cos\left(\theta\right)\right)\exp\left[\left(-iu\Phi+\alpha n\Phi q\cos\left(\theta_{q}\right)\right)\cos\left(\theta\right)\right]I_{0}\left(\alpha n\Phi q\left|\sin\left(\theta\right)\sin\left(\theta_{q}\right)\right|\right)}\\
{\displaystyle =4\pi\frac{\sin\left(\Phi\sqrt{\left(\mathbf{u}+i\alpha n\mathbf{q}\right)^{2}}\right)}{\Phi\sqrt{\left(\mathbf{u}+i\alpha n\mathbf{q}\right)^{2}}}.}\end{array}\label{eq:int_omega}\end{equation}
When we replace this expression into Eq.(\ref{eq:int_phi_q}) we obtain
the following form for $I_{\Phi}$

\begin{equation}
I_{\Phi}=4\pi\int_{0}^{\infty}d\Phi\sin\left(\Phi\right)\frac{\sin\left(\Phi\sqrt{\left(\mathbf{u}+i\alpha n\mathbf{q}\right)^{2}}\right)}{\sqrt{\left(\mathbf{u}+i\alpha n\mathbf{q}\right)^{2}}}\exp\left(-\frac{\alpha n}{2}(\Phi^{2}+q^{2})+i\mathbf{q}\cdot\mathbf{u}\right).\end{equation}
The solution of this integral is

\begin{equation}
{\displaystyle =4\pi\sqrt{\frac{\pi}{2\alpha n}}\frac{\sinh\left(\frac{1}{\alpha n}\sqrt{\left(\mathbf{u}+i\alpha n\mathbf{q}\right)^{2}}\right)}{\sqrt{\left(\mathbf{u}+i\alpha n\mathbf{q}\right)^{2}}}}\exp\left(-\frac{1}{2\alpha n}-\frac{\left(\mathbf{u}+i\alpha n\mathbf{q}\right)^{2}}{2\alpha n}{\displaystyle +i\mathbf{q}\cdot\mathbf{u}-\frac{\alpha n}{2}\mathbf{q}^{2}}\right).\label{eq:final_int_PHI}\end{equation}
Consequently, the characteristic function can be written as follows 

\begin{equation}
C(q)=\frac{2\pi}{N}\left[4\pi\sqrt{\frac{\pi}{2\alpha n}}\right]^{n}\int_{0}^{\infty}du\int_{-1}^{1}d\left(\cos\left(\theta\right)\right)\left[\frac{\sinh\left(\frac{1}{\alpha n}\sqrt{\left(\mathbf{u}+i\alpha n\mathbf{q}\right)^{2}}\right)}{\sqrt{\left(\mathbf{u}+i\alpha n\mathbf{q}\right)^{2}}}\right]^{n}u^{2}\exp\left(-\frac{u^{2}}{2\alpha}\right),\label{eq:charact_important}\end{equation}
where $N$ is the norm which is defined in such a way that $C\left(0\right)=1$. 

In the limit of $\alpha\rightarrow\infty$, Eq.(\ref{eq:charact_important})
approaches the asymptotic limit given by the formula

\begin{equation}
C_{\alpha\rightarrow\infty}(q)\sim\left[\frac{\sin\left(q\right)}{q}\right]^{n},\label{eq:alpla_infinity}\end{equation}
which is the characteristic function of the Random Flight Model.$^{28}$

On the other hand, if $\alpha\rightarrow0$,

\begin{equation}
\begin{array}{c}
{\displaystyle C_{\alpha\rightarrow0}(q)\sim\int_{0}^{\infty}du\int d\Omega u^{2-n}\exp\left(-\frac{u^{2}}{2\alpha}+iqn\cos\left(\theta\right)\right)\sinh^{n}\left(u/\alpha n\right)}\\
{\displaystyle \sim\frac{\sin\left(qn\right)}{qn},}\end{array}\label{eq:alpha_zero}\end{equation}
and the Rod-like limit is recovered.$^{29}$ 

An expression for the entire range of values of $\alpha$ is obtained
from Eq.(\ref{eq:charact_important}). The first step is to compute
the integral over the angles, that is

\begin{equation}
C_{\Omega}(q)=2\pi\int_{-1}^{1}d\left(\cos\left(\theta\right)\right)\left[\frac{\sinh\left(\frac{1}{\alpha n}\sqrt{\left(\mathbf{u}+i\alpha n\mathbf{q}\right)^{2}}\right)}{\sqrt{\left(\mathbf{u}+i\alpha n\mathbf{q}\right)^{2}}}\right]^{n},\label{eq:C_omega}\end{equation}

A variable transformation $v^{2}=\left(u^{2}-\left(\alpha nq\right)^{2}+2iu\alpha nq\cos\left(\theta\right)\right)$
allows us to rewrite $C_{\Omega}(q)$ as follows 

\begin{equation}
C_{\Omega}(q)=\frac{2\pi}{iqu}\left(\alpha n\right)^{1-n}\int_{-\frac{\left(u-i\alpha nq\right)}{\alpha n}}^{\frac{\left(u+i\alpha nq\right)}{\alpha n}}dv\frac{\sinh^{n}\left(v\right)}{v^{n-1}}.\label{eq:C_omega_middle}\end{equation}

Replacing the hyperbolic sine by exponentials and, after expanding
the resulting binomial, we can rewrite $C_{\Omega}(q)$ as follows 

\begin{equation}
C_{\Omega}(q)=\frac{\pi}{iqu}\left(2\alpha n\right)^{1-n}\sum_{k=0}^{n}\left(-1\right)^{k}\left(\begin{array}{c}
n\\
k\end{array}\right)\int_{-\frac{\left(u-i\alpha nq\right)}{\alpha n}}^{\frac{\left(u+i\alpha nq\right)}{\alpha n}}dv\frac{\exp\left[\left(n-2k\right)v\right]}{v^{n-1}},\label{eq:C_omega_Int}\end{equation}
which is easily calculated in terms of the Incomplete Gamma function.$^{27}$
The result is

\begin{equation}
C_{\Omega}(\mathbf{q})=-\frac{2\pi i}{qu}\left(2\alpha n\right)^{1-n}\sum_{k=0}^{n}\left(-1\right)^{k}\left(\begin{array}{c}
n\\
k\end{array}\right)\left(2k-n\right)^{n-2}Im\left\{ \Gamma\left(2-n,\left(2k-n\right)\frac{\left(u+i\alpha nq\right)}{\alpha n}\right)\right\} .\label{eq:C_omega_end}\end{equation}

Going back to $C(q)$, we note that the range of integration can be
expanded to all real values of $u$ by adding $C(q)$ to its complex
conjugate to get an integrand invariant under the transformation $u\rightarrow-u$.
Then, we integrate it by parts to get 

\begin{equation}
\begin{array}{c}
{\displaystyle C\left(q\right)=\frac{\alpha\pi^{\left(3n/2+1\right)}}{qN}\left(\frac{2}{\alpha n}\right)^{\left(n/2+1\right)}{\displaystyle \sum_{k=0}^{n}}\left(\begin{array}{c}
n\\
k\end{array}\right)\left(-\right)^{k}}\\
{\displaystyle Im\left\{ \int_{-\infty}^{+\infty}du\left(u+iq\alpha n\right)^{1-n}\exp\left[-\frac{u^{2}}{2\alpha}+\frac{n-2k}{\alpha n}u+iq\left(n-2k\right)\right]\right\} .}\end{array}\label{charac}\end{equation}

The calculation of the integral in Eq.(\ref{charac}) is straightforward.
The result is

\begin{equation}
\begin{array}{c}
{\displaystyle C\left(q\right)=\frac{4\pi^{\frac{3(n+1)}{2}}}{N\alpha^{n-1}n^{\frac{n}{2}+1}}\sum_{k=0}^{n}\left(\begin{array}{c}
n\\
k\end{array}\right)\left(-1\right)^{k}\exp\left(\frac{\left(n-2k\right)^{2}}{2\alpha n^{2}}\right)Re\left(\exp\left[iq\left(n-2k\right)-\frac{i\pi n}{2}\right]\right.}\\
{\displaystyle \left.U\left(\frac{n-1}{2},\frac{1}{2},-\left[\frac{n-2k}{\sqrt{2\alpha}n}+iqn\sqrt{\frac{\alpha}{2}}\right]^{2}\right)\right)},\end{array}\label{eq:C_deri}\end{equation}
where $U(a,b,c)$ is Kummer's function$^{27}$ which can be rewritten
in terms of Parabolic Cylinder, Whittaker or other functions. Its
mathematical properties like recurrence relations and integral representations
are very well known and tabulated.$^{27}$ But, although its mathematical
properties are many, we were unable to compute the real part in Eq.(\ref{eq:C_deri})
in a \emph{compact} mathematical form. 

Kummer's function has a cut on the negative real axis. Consequently,
the evaluation of the norm $N$ and the mean square end-to-end distance
$\left\langle \mathbf{R}^{2}\right\rangle $ cannot be obtained from
the Taylor expansion of Eq.(\ref{eq:C_deri}) in powers of the wave
vector $\mathbf{q}$. Taking into account that both real and imaginary
parts of $U(a,b,-\left|c\right|)$ have well-known expressions$^{30}$
we used Eq.(\ref{charac}). Firstly, we interchanged the sum with
the imaginary operator and carried out the sum which gives a hyperbolic
sine to the $n$th power. Afterward, the integrand was expanded in
powers of the wave vector to third order and the integrals evaluated.
The results are

\begin{equation}
N=\frac{\left(2\pi\right)^{3n/2+1}}{\left(\alpha n\right)^{3n/2-3}}I_{n-2}^{n},\label{eq:Norm}\end{equation}
and 

\begin{equation}
\left\langle \mathbf{R}^{2}\right\rangle =\frac{\left(2\pi\right)^{3n/2+1}}{N\left(\alpha n\right)^{3n/2-3}}\left\{ \left(n^{2}-n\right)I_{n-2}^{n-2}-n^{2}\left[2\alpha\left(n-1\right)-1\right]I_{n-2}^{n}-\left(n-1\right)\left(n-2\right)I_{n}^{n}\right\} ,\label{eq:Rsquare}\end{equation}

where

\[
I_{m}^{p}\equiv\int_{-\infty}^{\infty}dt\frac{\sinh^{p}\left(t\right)}{t^{m}}\exp\left[-\frac{\alpha n^{2}}{2}t^{2}\right]\]

\[
=\frac{\sqrt{\pi}n^{m-1}\alpha^{1/2-m/2}}{2^{p+m/2-1/2}}\left\{ -\cos\left(\frac{\pi p}{2}\right)\left(\begin{array}{c}
p\\
p/2\end{array}\right)\frac{\sqrt{\pi}\left(-\right)^{m/2}}{\Gamma\left(m/2+1/2\right)}\right.\]
\[
\left.+2\sum_{k=0}^{\left[p/2\right]}\exp\left[\frac{\left(p-2k\right)^{2}}{2\alpha n^{2}}\right]\left(\begin{array}{c}
p\\
k\end{array}\right)\left(-\right)^{k}Re\left\{ \left(-i\right)^{m}U\left(\frac{m}{2},\frac{1}{2},-\frac{\left(p-2k\right)^{2}}{2\alpha n^{2}}\right)\right\} \right\} .\]

\subsection{Evaluation of the Polymer Propagator }

We now proceed to evaluate the polymer propagator. For this purpose,
the starting point is the Fourier transform of the characteristic
function given by Eq.(\ref{charac}). Firstly, we interchange the
order of integration as follows 

\begin{equation}
\begin{array}{c}
{\displaystyle P\left(R\right)=R^{-1}N^{-1}\frac{\alpha}{2}\pi^{\left(3n/2-1\right)}\left(\frac{2}{\alpha n}\right)^{\left(n/2+1\right)}{\displaystyle \sum_{k=0}^{n}}\left(\begin{array}{c}
n\\
k\end{array}\right)\left(-\right)^{k}}\\
{\displaystyle \int_{-\infty}^{+\infty}du\exp\left[-\frac{u^{2}}{2\alpha}+\frac{n-2k}{\alpha n}t\right]Im\left\{ \int_{0}^{\infty}dq\sin\left(qR\right)\left(u+iq\alpha n\right)^{1-n}\exp\left[iq\left(n-2k\right)\right]\right\} ,}\end{array}\label{prop}\end{equation}
and evaluate the imaginary part of the $q$-integral assuming that
$n$ is even. After replacing $\sin\left(qR\right)$ by its definition
in terms of exponentials, the propagator can be written in terms of
the following integrals 

\begin{equation}
W^{_{-}^{+}}(R)=\frac{\left(-1\right)^{\frac{n}{2}}}{2\left(\alpha n\right)^{n-1}}Im\left\{ \int_{0}^{+\infty}dq\frac{\exp\left(iq\left(n-2k\pm R\right)\right)}{\left(q-iu/\alpha n\right)^{n-1}}\right\} .\label{eq:W+-}\end{equation}
which can be solved if we consider $q$ to be a complex variable and
apply Cauchy's theorem. 

The evaluation of the $W$-integrals requires special attention to
the location of the pole $q=iu/\alpha n$ because $u$ can be positive
or negative. For reasons of notation, we define $W_{1}^{+},W_{2}^{+},W_{1}^{-},W_{2}^{-}$
such that the subindices $1$ and $2$ indicate a solution valid for
$u\geq0$ and $u\leq0$, respectively. 

We start by evaluating $W^{+}$ using Cauchy's theorem. If $\left(n-2k+R\right)\geq0$,
we choose the contour of integration C1 given in Fig. 1. The result
for $u\geq0$ is

\begin{equation}
W_{1}^{+}\left(R\right)=-\frac{\pi}{2\left(\alpha n\right)^{n-1}\left(n-2\right)!}\left\{ \left(n-2k+R\right)^{n-2}\exp\left[-\frac{t}{\alpha n}\left(n-2k+R\right)\right]\right\} .\label{eq:W+1}\end{equation}
On the other hand, if $u\leq0$, then $W_{2}^{+}\left(R\right)=0$. 

The other possibility is $\left(n-2k+R\right)\leq0$. Then, the contour
of integration in Fig. 1 should be C2. This implies that $W_{1}^{+}\left(R\right)=0$
and

\begin{equation}
W_{2}^{+}\left(R\right)=\frac{\pi}{2\left(\alpha n\right)^{n-1}\left(n-2\right)!}\left\{ \left(n-2k+R\right)^{n-2}\exp\left[-\frac{t}{\alpha n}\left(n-2k+R\right)\right]\right\} .\label{eq:W+2}\end{equation}

The evaluation of $W^{-}$ is done in a similar way. If $\left(n-2k-R\right)\geq0$,
then the contour of integration should be C1 and the results are $W_{2}^{-}\left(R\right)=0$
and $W_{1}^{-}(R)=W_{1}^{+}(-R)$. Otherwise, if $\left(n-2k-R\right)\leq0$,
then the contour is C2 and the results are $W_{1}^{-}\left(R\right)=0$
and $W_{2}^{-}(R)=-W_{1}^{+}(-R)$. 

Substituting these results into Eq.(\ref{prop}) and evaluating the
following integrals

\begin{equation}
\begin{array}{c}
{\displaystyle {\displaystyle A_{k}\left(R\right)\equiv\left(n-2k+R\right)^{n-2}\int_{0}^{+\infty}dt\exp\left[-\frac{t^{2}}{2\alpha}-Rt/\alpha Ln\right]}}\\
{\displaystyle {\displaystyle =\left(n-2k+R\right)^{n-2}\sqrt{\pi\alpha/2}\exp\left[R^{2}/2\alpha n^{2}\right]\left[1-erf\left(\frac{R}{\sqrt{2\alpha}n}\right)\right],}}\end{array}\label{eq:gb}\end{equation}
and

\begin{equation}
\begin{array}{c}
{\displaystyle {\displaystyle B_{k}\left(R\right)\equiv\left(n-2k-R\right)^{n-2}\int_{-\infty}^{0}dt\exp\left[-\frac{t^{2}}{2\alpha}+Rt/\alpha n\right]}}\\
{\displaystyle {\displaystyle =\left(n-2k-R\right)^{n-2}\sqrt{\pi\alpha/2}\exp\left[R^{2}/2\alpha n^{2}\right]\left[1-erf\left(\frac{R}{\sqrt{2\alpha}n}\right)\right],}}\end{array}\label{eq:Bk}\end{equation}
 then, the polymer propagator given by Eq.(\ref{prop}) becomes\begin{equation}
{\displaystyle P\left(R\right)=\frac{\pi^{\left(3n/2+1/2\right)}\alpha^{3\left(1-n\right)/2}2^{n/2+1/2}}{n^{3n/2}\left(n-2\right)!N}\exp\left[{\displaystyle \frac{R^{2}}{2\alpha n^{2}}}\right]{\displaystyle \sum_{k=0}^{\left[\left(n-R\right)/2\right]}}\left(\begin{array}{c}
n\\
k\end{array}\right)\left(-\right)^{k}\frac{\left(n-2k-R\right)^{n-2}}{R}}.\label{eq:propsum}\end{equation}
This expression, which was obtained assuming that $n$ was even, is
easily extended to any number of segments by analytic continuation.

A comparison of this result with the one of the Random Flight Model$^{23}$
shows that we can write the final expression for the propagator in
a very simple and insightful way 

\begin{equation}
P\left(R\right)=\frac{\alpha^{3/2}\left(2\pi\right)^{\left(3n/2+3/2\right)}}{N\left(\alpha n\right)^{3n/2}}P_{RFM}\left(R\right)\exp\left[\frac{R^{2}}{2\alpha n^{2}}\right],\label{eq:propend}\end{equation}
where the RMF stands for Random Flight Model.

\section{Results and discussion}

The polymer propagator described by Eq.(\ref{eq:propend}) clearly
shows that our treatment of the KP model captures the limit of fully
flexible chains $\left(\alpha\rightarrow\infty\right)$ exactly. Indeed,
in this limit, Eq.(\ref{eq:propend}) approaches the well-established
expression of the Random Flight Model. Furthermore, Eq.(\ref{eq:propend})
clearly shows how the stiffness of the polymer backbone modifies the
propagator of the RFM. In the other limit $\left(\alpha\rightarrow0\right)$,
the exponential grows. Consequently, those configurations of the polymer
chain with large end-to-end distance have higher probability of been
realized than the configurations with small end-to-end distance. This
is the correct physical behavior and is a consequence of the free
energy penalty related to the formation of hairpins. But, although
the exponential function grows in this limit, the propagator of the
RFM puts an upper limit to the possible values of the end-to-end distance.
This limit is the number of segments of the chain. This is clearly
shown in the upper limit of the sum in Eq.(\ref{eq:propsum}). Thus,
the polymer chain satisfies the local inextensibility constraint.
Another very important consequence of Eq.(\ref{eq:propend}) is its
mathematical simplicity which makes this expression of the polymer
propagator a very good starting point for a perturbative treatment
of the KP model as mentioned before. 

We plot the normalized radial distribution as a function of the end-to-end
distance in Figs. 2, 3, 4 and 5. The values of the parameters chosen
are: $n=$6, 10, 20 and 30 Kuhn segments, and $\alpha=$0.01, 0.02,
0.03, 0.05, 0.1 and 0.75. The figures clearly show that the location
of the peak in the polymer propagator (multiplied by $R^{2}$) moves
toward larger values of $R$ when the stiffness of the polymer backbone
increases. This behavior is in good qualitative agreement with previous
results arising from computer simulation studies$^{19}$ and theoretical
approaches based on the Maximum Entropy Principle.$^{20}$ This is
the correct result because the stiffer the polymer backbone, the higher
the energetic penalty to bend the chain. Consequently, those configurations
of the macromolecule with small end-to-end distance will be more and
more hindered as the stiffness increases, while those configurations
with large end-to-end distance should be more and more favored. Therefore,
the peak should shift toward larger values of $R$ when the stiffness
increases as shown by the figures.

Figures 6 and 7 show the polymer propagator for polymer chains with
6, 10, 20 and 30 Kuhn segments and a fixed value of the semiflexibility
parameter. Theses figures show that the longer the polymer is, the
more it behaves like a flexible chain since the location of peak (=end-to-end
distance divided by the number of segments) moves toward smaller values.
In other words, the longer the polymer is, the less relevant the stiffness
of the backbone becomes. 

The two aforementioned effects can be easily rationalized in terms
of Eq.(\ref{eq:propend}). Observe that the stiffness parameter $\alpha$
and the length of the chain $n$ do not appear as independent variables
but, as the product $\alpha n^{2}$. Therefore, making the chain longer
is equivalent to increasing the semiflexibility parameter thus making
the polymer more flexible. 

Another interesting result is the limit of very long chains. It is
known that in this limit the polymer propagator of the RFM approaches
the one of the Gaussian Chain Model.$^{28}$ Thus, Eq.(\ref{eq:propend})
should approach the following expression

\begin{equation}
P\left(R\right)\sim\exp\left[-\frac{3R^{2}}{2n}+\frac{R^{2}}{2\alpha n^{2}}\right]=\exp\left[-\frac{3R^{2}}{2n}\left(1-\frac{1}{3\alpha n}\right)\right]\equiv\exp\left[-\frac{3R^{2}}{2n\left(l_{K}^{e}\right)^{2}}\right],\label{eq:longprop}\end{equation}
where we have defined an {}``effective'' Kuhn length $l_{K}^{e}$
(in units of the bare one, $l_{K}$) as follows

\begin{equation}
\left(l_{K}^{e}\right)^{2}=\frac{1}{\left(1-\frac{1}{3\alpha n}\right)}.\label{eq:renorma}\end{equation}
Observe that this renormalized Kuhn length approaches the bare one
in the limits of very long , $n\rightarrow\infty$, or very flexible
chains, $\alpha\rightarrow\infty$. This is the expected result and
can be rationalized using the arguments presented in the previous
paragraph. In addition, note that as  the chain becomes shorter or
stiffer, Eq.(\ref{eq:renorma}) shows that the effective Kuhn length
grows. Therefore, the effects of semiflexibility become more relevant. 

Figures 8 and 9 show the behavior of $C\left(q\right)$, Eq.(\ref{eq:C_deri}),
as a function of the wave vector $q$ for different values of $\alpha$
and two values of $n$ (10 and 20). The figures clearly show that
our treatment of the KP model predicts a smooth transition from the
rod-like to the flexible limit. Moreover, our results capture the
continuous change in the qualitative behavior of this function which
changes from a monotonically decreasing function in the flexible limit
to an oscillating function in the rod-like regime. In addition, our
computations predict that the decrease of the characteristic function
for small values of $q$ should be faster for stiff polymers than
for flexible ones. This is a consequence of the fact that rigid polymers
have a larger mean squared end-to-end distance than flexible ones
for a fixed chain length. 

The effect of chain length on the characteristic function is shown
in Fig. 10 for the particular cases of chains with 6, 10, 20 and 30
Kuhn segments, and a fixed value of the semiflexibility parameter
$\alpha(=0.01)$. This figure shows that the longer the polymer is,
the more it behaves like a flexible chain since the characteristic
function approaches the one of a flexible polymer chain. 

Figures 11 and 12 show the mean squared end-to-end distance $\left\langle \mathbf{R}^{2}\right\rangle $
as function of the semiflexibility parameter $\alpha$ for different
chain length. Specifically, Fig. 11 compares our approximate solution,
Eq.(\ref{eq:Rsquare}), for $n=8$ with the exact one.$^{5}$ This
figure shows that our approximate solution captures the limits of
flexible and rigid polymers exactly and provides an very good behavior
in the crossover. Figure 12 shows the behavior of $\left\langle \mathbf{R}^{2}\right\rangle $
as a function of $\alpha$ for chains with 5, 8, and 10 Kuhn segments. 

Finally, we compute the exponent $2\nu$ in the relationship $\left\langle \mathbf{R}^{2}\right\rangle =n^{2\nu}$.
Figure 13 shows the plot of $ln\left\{ \left\langle \mathbf{R}^{2}\right\rangle \right\} $
as function of the number of segments, $ln\left(n\right)$, for two
values of the semiflexibility parameter, $\alpha=0.001$ (rigid) and
$\alpha=0.75$ (flexible). The numerical values of the slopes are
$1.95$ and $0.98$, respectively. These results are in excellent
agreement with the expected values for the exponents of rodlike and
fully flexible chains. Thus, our approach also captures the fractal
dimensions of both limiting behaviors correctly. We have also studied
intermediate values of the semiflexibility parameter and observed
a smooth crossover from the rod-like to the flexible limit.

\section{conclusions}

In this paper we have proposed a new treatment of the KP model based
on a generating function. The advantage of our approach is two fold.
Firstly, the evaluation of the most relevant statistical properties
of the model is straight forward since they can be obtained as derivatives
or integrals of the generating function. This makes the evaluation
of the generating function the crucial step for the solution of the
KP model. In our treatment of this model, we were able to devise one
approximation that was able to capture the flexible and rigid limits
exactly and, moreover, was able to provide a smooth crossover behavior
between the two aforementioned limiting regimes. This crossover has
all the correct qualitative features as discussed in the previous
section. Furthermore, this approximation respects the local, not global,
inextensibility constraint. Secondly, our treatment of the KP model
was able to provide a good, though not unique, {}``ground'' state
around which a perturbative treatment can be developed. We speculate
that this perturbative calculation should be able to correct the consequences
of our approximation in a systematic and controlled fashion. In other
words, this perturbative analysis should account for the corrections
arising from the neglected non-local terms in the kernel and deviations
of the diagonal terms from the pre-averaging approximation used. 

The results presented in the paper are valid for polymers with any
number of segments and value of the stiffness of the polymer backbone.
Specifically, we computed the characteristic function, the mean squared
end-to-end distance and the polymer propagator of the KP model. Other
averages like the radius of gyration and/or correlation functions
like the single chain structure factor or tangent-tangent correlation
functions can be computed in a straightforward manner, too.

It is worth mentioning that the behaviors predicted for the characteristic
function, polymer propagator and mean squared end-to-end distance
are in very good qualitative and quantitative agreement with results
arising from other studies of the KP and Worm-like Chain Model. This
agreement makes us speculate that our future evaluation of the structure
factor, the radius of gyration and other correlation functions calculated
within the present level of approximation should give good results
at least at the qualitative level. 

Perhaps, the most important contribution of our work is the derived
polymer propagator. The mathematical expression obtained for the propagator
is very simple, compact and insightful. It becomes the exact expression
of the Random Flight Model in the flexible limit. But, as the stiffness
of the polymer backbone increases, the propagator of the RFM is modified
by an exponential which depends on the semiflexibility parameter.
As a consequence, the peak of the radial distribution function shifts
toward larger values of the end-to-end distance when the polymer becomes
stiffer. In the limit of infinitely stiff polymers, the propagator
becomes a delta function centered at the total contour length of the
chain. This is the correct result because the stiffer the polymer
backbone, the higher the energetic penalty to bend the chain. This
shift of the peak is accompanied by an increase in its height and
a decrease in the width of the distribution.

\section{Acknowledgments}

We acknowledge the National Science Foundation, Grant \# CHE-0132278
(CAREER), the Ohio Board of Regents Action Fund, Proposal \# R566
and The University of Akron for financial support.

\appendix

\section{Some Mathematical Aspects of the Evaluation of the Generating Function}

In this appendix we explain how the final expression of the generating
function, Eq.(\ref{eq:Final_C}), was obtained. For the purpose of
clarity we define

\begin{equation}
\Upsilon_{k}\equiv\left(\Phi_{k}+i\Psi_{kjp}\right),\label{eq:AA1}\end{equation}
then, the $u_{k}$-integrals in Eq.(\ref{eq:stratanovich}) can be
evaluated exactly using saddle point which is the solution to the
following equations

\begin{equation}
\mathbf{u}_{1}-\mathbf{u}_{0}=-\kappa\Upsilon_{0},\label{eq:Acontitions1}\end{equation}

\begin{equation}
\mathbf{u}_{n-1}-\mathbf{u}_{n-2}=\kappa\Upsilon_{n-1},\label{eq:Aconditions2}\end{equation}

\begin{equation}
\mathbf{u}_{k+1}-2\mathbf{u}_{k}+\mathbf{u}_{k-1}=-\kappa\Upsilon_{k},\:\:\:\;\;\; k=1..n-2,\label{eq:Aconditions3}\end{equation}
where $\kappa$ was defined in Eq.(\ref{eq:Kernel}). The solutions
to these equations are given by Eq.(\ref{eq:saddle_point}). Moreover,
the constraint for the sources, Eq.(\ref{eq:constrainsource}), is
a consequence of these equations. Therefore, if this constraint is
not satisfied, then Eq. (\ref{eq:saddle_point}) is not the saddle
point solution. In addition, those terms appearing in Eq.(\ref{eq:stratanovich})
can be rewritten as follows 

\begin{equation}
-\frac{1}{2\kappa}\sum_{k=0}^{n-2}\left(\mathbf{u}_{k+1}-\mathbf{u}_{k}\right)^{2}=-\frac{\kappa}{2}\sum_{k=0}^{n-2}\left(\sum_{m=0}^{k}\Upsilon_{m}\right)^{2},\label{eq:AA2}\end{equation}
and also

\begin{equation}
\sum_{k=0}^{n-1}\mathbf{u}_{k}\cdot\Upsilon_{k}=-\frac{\kappa}{2}\sum_{k=0}^{n-1}\sum_{m=0}^{n-1}\Upsilon_{m}\left|k-m\right|\Upsilon_{k}.\label{eq:AA3}\end{equation}
 We can simplify the sum of these two terms further if we note that

\[
\sum_{k=0}^{n-1}\sum_{m=0}^{n-1}\Upsilon_{m}\left|k-m\right|\Upsilon_{k}+\sum_{k=0}^{n-2}\left(\sum_{m=0}^{k}\Upsilon_{m}\right)^{2}\]

\begin{equation}
=\sum_{k=0}^{n-1}\sum_{m=0}^{n-1}\Upsilon_{m}\left|k-m\right|\Upsilon_{k}+\sum_{k=0}^{n-2}\sum_{m=0}^{k}\Upsilon_{m}\Upsilon_{k}\left(n-1-k\right)+\sum_{k=0}^{n-2}\sum_{m=k+1}^{n-1}\Upsilon_{m}\Upsilon_{k}\left(n-1-m\right)\label{eq:AA4}\end{equation}

\[
=\sum_{k=0}^{n-1}\sum_{m=0}^{n-1}\Upsilon_{m}\left|k-m\right|\Upsilon_{k}-\sum_{k=0}^{n-2}\sum_{m=k+1}^{n-1}\Upsilon_{m}\left(m-k\right)\Upsilon_{k}+\left(n-1\right)\sum_{k=0}^{n-2}\sum_{m=0}^{n-1}\Upsilon_{m}\Upsilon_{k}\]
where the last double sum vanishes because of the constraint, Eq.(\ref{eq:constrainsource}).
After changing the order of summation we arrive at

\begin{equation}
\sum_{k=0}^{n-1}\sum_{m=0}^{n-1}\Upsilon_{m}\left|k-m\right|\Upsilon_{k}-\sum_{k=1}^{n-1}\sum_{m=0}^{k-1}\Upsilon_{m}\left(k-m\right)\Upsilon_{k}.\label{eq:AA5}\end{equation}

The first double sum, apart from the term $k=0$, can be separated
into two terms. One for $m\leq k-1$ and another one for $m\geq k$
such that the former cancels the second double sum in the previous
expression. As a result we obtain

\begin{equation}
\begin{array}{c}
{\displaystyle \sum_{k=0}^{n-1}\Upsilon_{0}k\Upsilon_{k}+\sum_{k=1}^{n-1}\sum_{m=k}^{n-1}\Upsilon_{m}\left|k-m\right|\Upsilon_{k}}\\
{\displaystyle =\sum_{k=0}^{n-1}\sum_{m=0}^{n-1}\Upsilon_{m}(m-k)\Theta(m-k)\Upsilon_{k}.}\end{array}\label{eq:AA6}\end{equation}

If we now use the constraint for the $\Upsilon_{k}$'s, Eq.(\ref{eq:constrainsource}),
we can readily get the quadratic term in Eq.(\ref{eq:Final_C}).

\subsection*{References}

\noindent $^{1}${\small G. Porod,} \textit{\small Monatsh. Chem.}
\textbf{\small 80}{\small , 251 (1949); O. Kratky and G. Porod,} \textit{\small Rec.
Trav. Chim.} \textbf{\small 68}{\small , 1106 (1949). }{\small \par}

\noindent {\small $^{2}$N. Saito, K. Takahashi and Y. Yunoki,} \emph{\small J.
Phys. Soc. Japan}{\small ,} \textbf{\small 22}{\small , 219 (1967). }{\small \par}

\noindent {\small $^{3}$H. Yamakawa,} \textit{\small Helical Worm-like
Chains in Polymer Solutions} {\small (Springer, Berlin, 1997). }{\small \par}

\noindent {\small $^{4}$H. E. Daniels,} \textit{\small Proc. Roy.
Soc. Edinburgh} \textbf{\small A63}{\small , 290 (1952).}{\small \par}

\noindent {\small $^{5}$H. Benoit and P. Doty,} \textit{\small J.
Phys. Chem.} \textbf{\small 57}{\small , 958 (1953). }{\small \par}

\noindent {\small $^{6}$M. Fixman and J. Kovac,} \textit{\small J.
Chem. Phys.} \textbf{\small 58}{\small , 1564(1973). }{\small \par}

\noindent {\small $^{7}$R. A. Harris and J. E. Hearst,} \textit{\small J.
Chem. Phys.} \textbf{\small 44}{\small , 2595 (1966).}{\small \par}

\noindent {\small $^{8}$W. Gobush, H. Yamakawa, W. H. Stockmayer
and W. S. Magee,} \textit{\small J. Chem. Phys.} \textbf{\small 57}{\small ,
2839 (1972).}{\small \par}

\noindent {\small $^{9}$H. Yamakawa and W. H. Stockmayer,} \textit{\small J.
Chem. Phys.} \textbf{\small 57}{\small , 2843 (1972).}{\small \par}

\noindent {\small $^{10}$K. F. Freed in} \textit{\small Advances
in Chemical Physics}{\small , Vol. 22, Pages 1-128, I. Prigogine and
Stuart A. Rice Editors (John Wiley and Sons, New York, 1972). }{\small \par}

\noindent {\small $^{11}$M. G. Bawendi and K. F. Freed,} \textit{\small J.
Chem. Phys.} \textbf{\small 83}{\small , 2491 (1985).}{\small \par}

\noindent {\small $^{12}$J. B. Lagowski, J. Noolandi and B. Nickel,}
\textit{\small J. Chem. Phys.} \textbf{\small 95}{\small , 1266 (1991).}{\small \par}

\noindent {\small $^{13}$R. G. Winkler, P. Reineker and L. Harnau,}
\textit{\small J. Chem. Phys.} \textbf{\small 101}{\small , 8119 (1994). }{\small \par}

\noindent {\small $^{14}$S. R. Zhao, C. P. Sun and W. X. Zhang,}
\textit{\small J. Chem. Phys.} \textbf{\small 106}{\small , 2520 (1997).}{\small \par}

\noindent {\small $^{15}$S. M. Bhattacharjee and M. Muthukumar,}
\textit{\small J. Chem. Phys.} \textbf{\small 86}{\small , 411(1987). }{\small \par}

\noindent {\small $^{16}$M. G. Poirier} \textit{\small et al.}{\small ,}
\textit{\small Phys. Rev. Lett.} {\small 86,} \textbf{\small 360}
{\small (2001); J. F. Leger} \textit{\small et at.}{\small ,} \textit{\small Phys.
Rev. Lett.} \textbf{\small 83}{\small , 1066 (1999); V. Parpura and
J. M. Fernandez,} \textit{\small Biophys. J.} \textbf{\small 71}{\small ,
2356 (1996); M. D. Wang} \textit{\small et al.}{\small ,} \textit{\small Biophys.
J.} \textbf{\small 72}{\small , 1335 (1997); A. D. Mehta, K. A. Pullen
and A. Spudich,} \textit{\small FEBS Lett.} \textbf{\small 430}{\small ,
23 (1998); A. D. Mehta, M. Rief and J. A. Spudich,} \textit{\small J.
Biol. Chem.} \textbf{\small 274}{\small , 14517 (1999),T. E. Fisher,
P. E. Marszalek and J. M. Fernandez,} \textit{\small Nat. Struct.
Biol.} \textbf{\small 7}{\small , 719 (2000); T. E. Fisher} \textit{\small et
al.}{\small ,} \textit{\small Trends Biochem. Sci.} \textbf{\small 24}{\small ,
379 (1999); M. Carrion-Vazquez} \textit{\small et al.}{\small ,} \textit{\small Prog.
Biophys. Mol. Biol.} \textbf{\small 74}{\small , 63 (2000); T. T.
Perkins, S. R. Quake, D. E. Smith and S. Chu,} \textit{\small Science}
\textbf{\small 264}{\small , 822 (1994); S. B. Smith, Y. Cui and C.
Bustamante,} \textit{\small Science} \textbf{\small 271}{\small ,
795 (1996); T. T. Perkins, D. E. Smith and S. Chu,} \textit{\small Science}
\textbf{\small 276}{\small , 2016 (1997); A. F. Oberhauser, P. E.
Marszalek, H. P. Erickson and J. M. Fernandez,} \textit{\small Nature
(London)} \textbf{\small 393}{\small , 181 (1998); F. Oesterhelt,
M. Rief and H. E. Gaub,} \textit{\small New J. Phys.} \textbf{\small 1}{\small ,
6.1 (1999); H. B. Li, W. K. Zhang, W. Q. Xu and X. Zhang,} \textit{\small Macromolecules}
\textbf{\small 33}{\small , 465 (2000). }{\small \par}

\noindent {\small $^{17}$J. F. Marko and E. D. Siggia,} \textit{\small Macromolecules}
\textbf{\small 28}{\small , 8759 (1995); K. Kroy and E. Frey,} \textit{\small Phys.
Rev. Lett.} \textbf{\small 77}{\small , 306 (1996). }{\small \par}

\noindent {\small $^{18}$P. L. Hansen and R. Podgornik,} \textit{\small J.
Chem. Phys.} \textbf{\small 114}{\small , 8637 (2001). }{\small \par}

\noindent {\small $^{19}$J. Wilhelm and E. Frey,} \textit{\small Phys.
Rev. Lett.} \textbf{\small 77}{\small , 2581 (1996).}{\small \par}

\noindent {\small $^{20}$R. G. Winkler,} \textit{\small J. Chem.
Phys.} \textbf{\small 118}{\small , 2919 (2003). }{\small \par}

\noindent {\small $^{21}$A. L. Kholodenko, Ann. Phys. 202, 186 (1990);
A. L. Kholodenko J. Chem. Soc. Faraday Trans. 91, 2473 (1995); A.
L. Kholodenko, Phys. Lett. A 141, 351 (1989); A. L. Kholodenko, Phys.
Lett. A 178, 180 (1993); A. Kholodenko and T. Vilgis Phys. Rev. E
50, 1257 (1994), A. L. Kholodenko and T. A. Vilgis Phys. Rev. E 52,
3973 (1995); A. L. Kholodenko, Macromolecules 26, 4179 (1993); A.
Kholodenko, M. Ballauff and M. Aguero Granados, Physica A 260, 267
(1998). }{\small \par}

\noindent {\small $^{22}$K. F. Freed,} \textit{\small Renormalization
Group Theory of Macromolecules} {\small (John Wiley \& Sons, New York,
1987); H. Fujita,} \textit{\small Polymer Solutions} {\small (Elsevier,
Amsterdam, 1990); M. Muthukumar and B. G. Nickel,} \textit{\small J.
Chem. Phys.} \textbf{\small 80}{\small , 5839 (1984);} \textit{\small ibid.}
\textbf{\small 86}{\small , 460 (1987).}{\small \par}

\noindent {\small $^{23}$M. C. Wang and E. Guth,} \textit{\small J.
Chem. Phys.} \textbf{\small 20}{\small , 1144 (1952); C. Hsiung, H.
Hsiung and A. Gordus,} \textit{\small J. Chem. Phys.} \textbf{\small 34}{\small ,
535 (1961); M. Marucho and G. A. Carri,} \textit{\small J. Math. Phys.}
{\small in press.}{\small \par}

\noindent {\small $^{24}$I.S. Gradshteyn and I.M. Ryzhik,} \emph{\small Table
of Integrals, Series, and Products} {\small (Academic Press, New York,
2000). }{\small \par}

\noindent {\small $^{25}$G. Arfken,} \textit{\small Mathematical
Methods for Physicists} {\small (Academic Press, New York, 1985).}{\small \par}

\noindent {\small $^{26}$J. Hubbard,} \emph{\small Phys. Rev. Lett.}
\textbf{\emph{\small 3}}\emph{\small ,} {\small 77 (1959); R.D. Stratonovich,}
\emph{\small Soviet Phys. Kokl.} \textbf{\small 2}{\small , 416 (1958).}{\small \par}

\noindent {\small $^{27}$M. Abramowitz and I. Stegun,} \emph{\small Handbook
of Mathematical Functions} {\small (Dover, New York, 1970). }{\small \par}

\noindent {\small $^{28}$H. Yamakawa, Modern Theory of Polymer Solutions
(Harper\&Row, New York, 1971).}{\small \par}

\noindent {\small $^{29}$J. S. Higgins and H. C. Benoit,} \textit{\small Polymers
and Neutron Scattering} {\small (Clarendon Press, Oxford, 1996).}{\small \par}

\noindent {\small $^{30}$H. Bateman,} \emph{\small Higher Transcendental
Functions}{\small , vol. 1 (Mc Graw Hill, New York, 1953) .}{\small \par}

\clearpage

\subsection*{List of Figures}

\begin{itemize}
\item FIG. 1. Contours of integration for the computation of the polymer
propagator.
\item FIG.~2: Normalized polymer propagator $4\pi R^{2}P\left(R\right)$
versus $R/n$ for $n=6$. Continuous line $\left(\alpha=0.01\right)$
, dashed line $\left(\alpha=0.02\right)$, dotted line $\left(\alpha=0.03\right)$,
dashed-dotted line $\left(\alpha=0.05\right)$, dashed-dotted-dotted
line $\left(\alpha=0.1\right)$ and dashed-dashed-dotted line $\left(\alpha=0.75\right)$.
\item FIG.~3: Normalized polymer propagator $4\pi R^{2}P\left(R\right)$
versus $R/n$ for $n=10$. Continuous line $\left(\alpha=0.01\right)$
, dashed line $\left(\alpha=0.02\right)$, dotted line $\left(\alpha=0.03\right)$,
dashed-dotted line $\left(\alpha=0.05\right)$, dashed-dotted-dotted
line $\left(\alpha=0.1\right)$ and dashed-dashed-dotted line $\left(\alpha=0.75\right)$.
\item FIG.~4: Normalized polymer propagator $4\pi R^{2}P\left(R\right)$
versus $R/n$ for $n=20$. Continuous line $\left(\alpha=0.01\right)$
, dashed line $\left(\alpha=0.02\right)$, dotted line $\left(\alpha=0.03\right)$,
dashed-dotted line $\left(\alpha=0.05\right)$, dashed-dotted-dotted
line $\left(\alpha=0.1\right)$ and dashed-dashed-dotted line $\left(\alpha=0.75\right)$.
\item FIG.~5: Normalized polymer propagator $4\pi R^{2}P\left(R\right)$
versus $R/n$ for $n=30$. Continuous line $\left(\alpha=0.01\right)$
, dashed line $\left(\alpha=0.02\right)$, dotted line $\left(\alpha=0.03\right)$,
dashed-dotted line $\left(\alpha=0.05\right)$, dashed-dotted-dotted
line $\left(\alpha=0.1\right)$ and dashed-dashed-dotted line $\left(\alpha=0.75\right)$.
\item FIG.~6: Normalized polymer propagator $4\pi R^{2}P\left(R\right)$
versus $R/n$ for $\alpha=0.01$. Continuous line $\left(n=6\right)$,
dotted line $\left(n=10\right)$, dashed line $\left(n=20\right)$
and dashed-dashed-dotted line $\left(n=30\right)$.
\item FIG.~7: Normalized polymer propagator $4\pi R^{2}P\left(R\right)$
versus $R/n$ for $\alpha=0.1$. Continuous line $\left(n=6\right)$,
dotted line $\left(n=10\right)$, dashed line $\left(n=20\right)$
and dashed-dashed-dotted line $\left(n=30\right)$.
\item FIG.~8: .Characteristic function $C\left(q\right)$ versus wave vector
$q$ for $n=10$. Dashed-dotted-dotted line $\left(\alpha=0\right)$
(the exact solution of rigid Model), dashed-dashed-dotted line $\left(\alpha=0.01\right)$,
long dashed line $\left(\alpha=0.04\right)$, dashed-dotted line $\left(\alpha=0.07\right)$,
dashed line $\left(\alpha=0.1\right)$, dotted line $\left(\alpha=0.75\right)$
and continuous line $\left(\alpha=\infty\right)$ (the exact solution
of the Random Flight Model).
\item FIG.~9: Characteristic function $C\left(q\right)$ versus wave vector
$q$ for $n=20$. Dashed-dotted-dotted line $\left(\alpha=0\right)$
(the exact solution of rigid Model), dashed-dashed-dotted line $\left(\alpha=0.01\right)$,
long dashed line $\left(\alpha=0.04\right)$, dashed-dotted line $\left(\alpha=0.07\right)$,
dashed line $\left(\alpha=0.1\right)$, dotted line $\left(\alpha=0.75\right)$
and continuous line $\left(\alpha=\infty\right)$ (the exact solution
of the Random Flight Model).
\item FIG.~10: Characteristic function $C\left(q\right)$ versus wave vector
$q$ for $\alpha=0.01$. Dotted line $\left(n=6\right)$, dashed line
$\left(n=10\right)$, dashed-dotted line $\left(n=20\right)$ and
continuous line $\left(n=30\right)$.
\item FIG.~11: Mean squared end-to-end distance $\left\langle \mathbf{R}^{2}\right\rangle $
versus the parameter $\alpha$ for $n=8$. The dotted line is our
approximate solution and the dashed line is the exact solution of
the KP model\cite{Benoit}.
\item FIG.~12: Mean square end-to-end distance $\left\langle \mathbf{R}^{2}\right\rangle $
(in logarithmic scale) versus the parameter $\alpha$. Dashed-dotted
line $\left(n=5\right)$, dashed line $\left(n=8\right)$ and dotted
line $\left(n=10\right)$.
\item FIG.~13: $ln\left\{ \left\langle \mathbf{R}^{2}\right\rangle \right\} $
versus the number of segments $ln\left(n\right)$. Continuous line
$(\alpha=0.001)$ and dashed line $(\alpha=0.75)$.
\end{itemize}
\clearpage

Comment: Figure 1, First Author: Marcelo Marucho, JCP

\begin{figure}
\includegraphics[%
  width=4in,
  keepaspectratio]{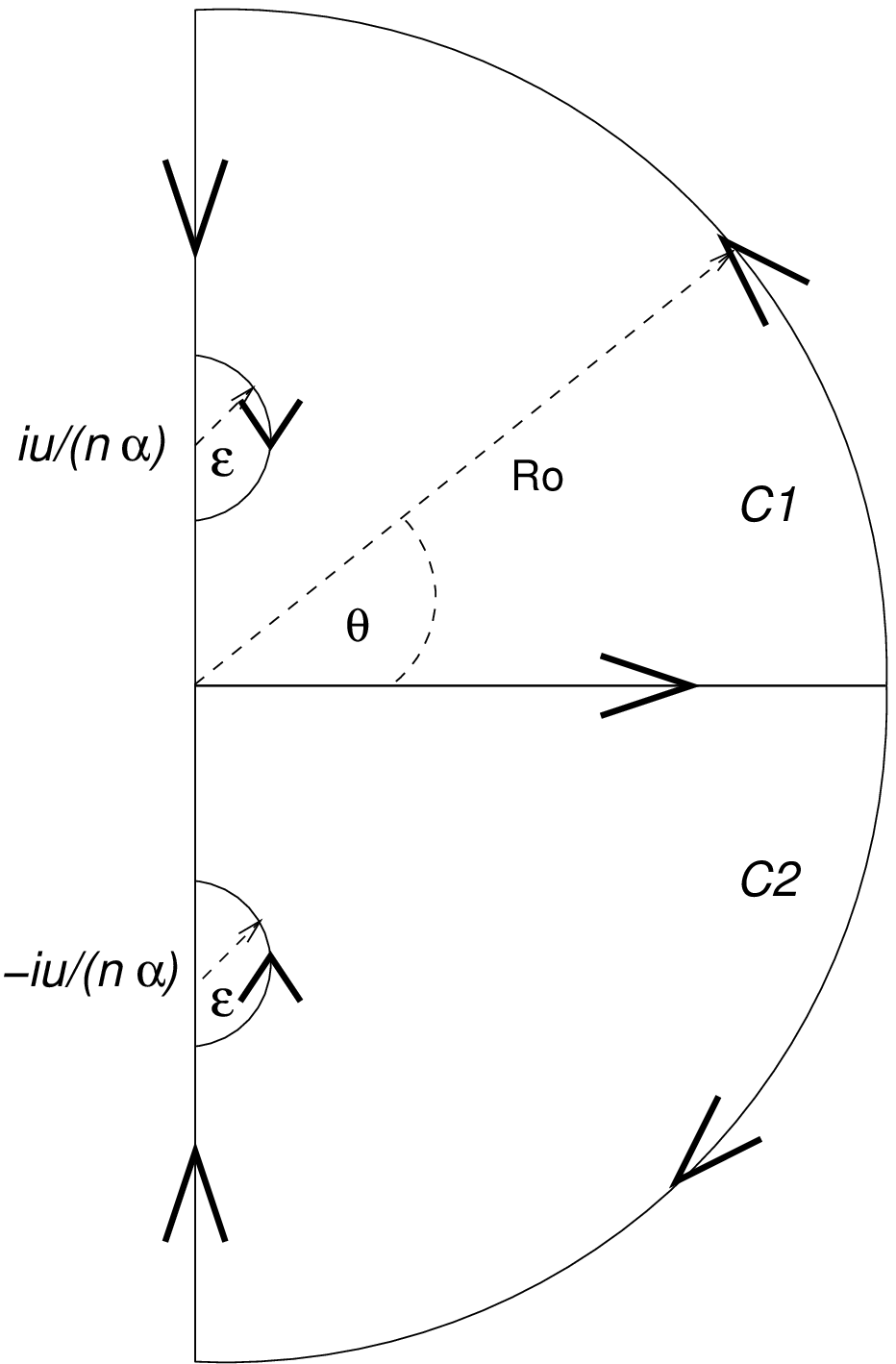}

\begin{flushleft}FIG. 1. Contours of integration for the computation
of the polymer propagator.\end{flushleft}
\end{figure}

\clearpage

Comment: Figure 2, First Author: Marcelo Marucho, JCP

\begin{figure}
\includegraphics[%
  width=4in,
  keepaspectratio,
  angle=270]{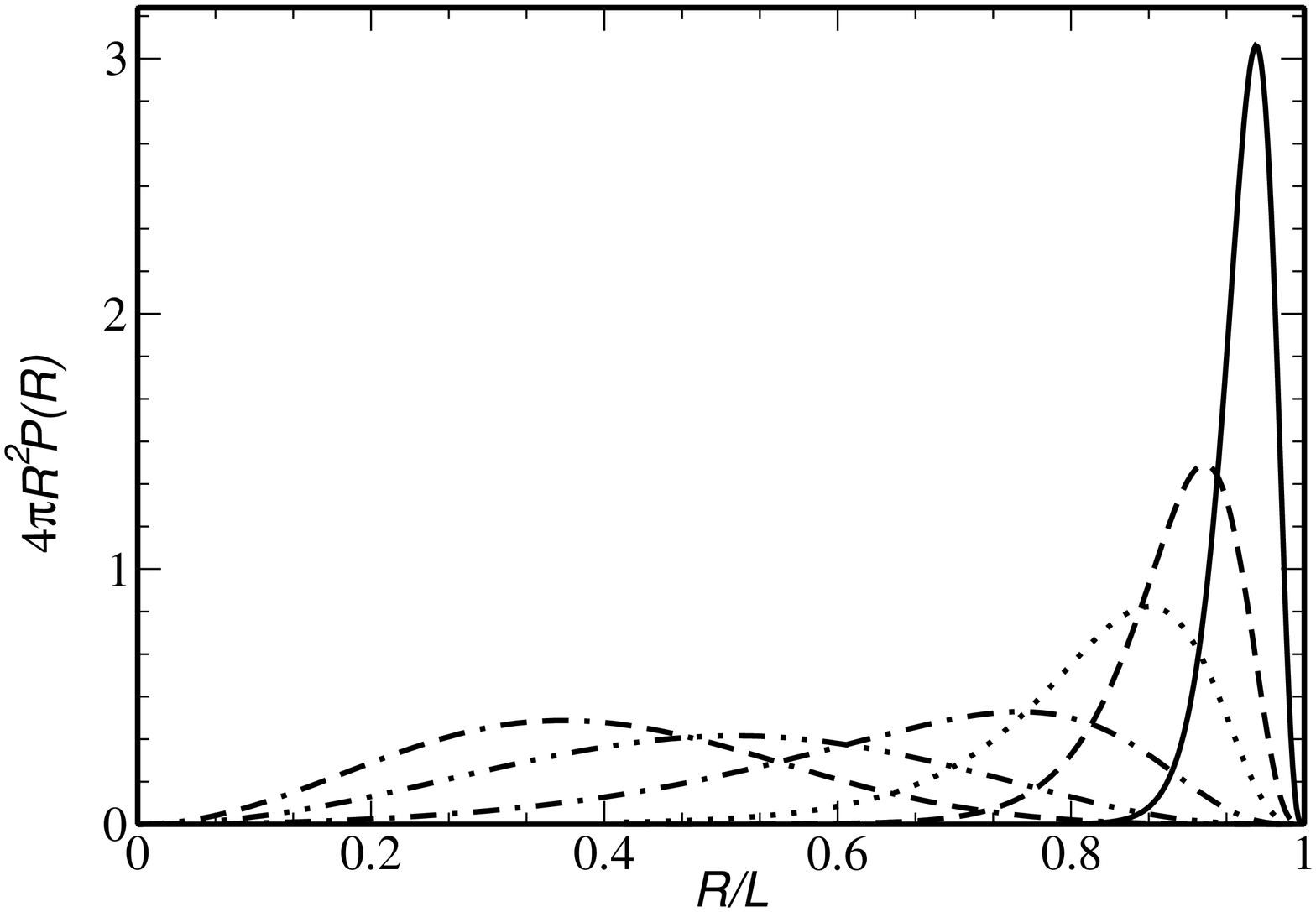}

\begin{flushleft}FIG. 2. Normalized polymer propagator $4\pi R^{2}P\left(R\right)$
versus $R/n$ for $n=6$. Continuous line $\left(\alpha=0.01\right)$
, dashed line $\left(\alpha=0.02\right)$, dotted line $\left(\alpha=0.03\right)$,
dashed-dotted line $\left(\alpha=0.05\right)$, dashed-dotted-dotted
line $\left(\alpha=0.1\right)$ and dashed-dashed-dotted line $\left(\alpha=0.75\right)$.\end{flushleft}
\end{figure}

\clearpage

Comment: Figure 3, First Author: Marcelo Marucho, JCP

\begin{figure}
\includegraphics[%
  width=4in,
  keepaspectratio,
  angle=270]{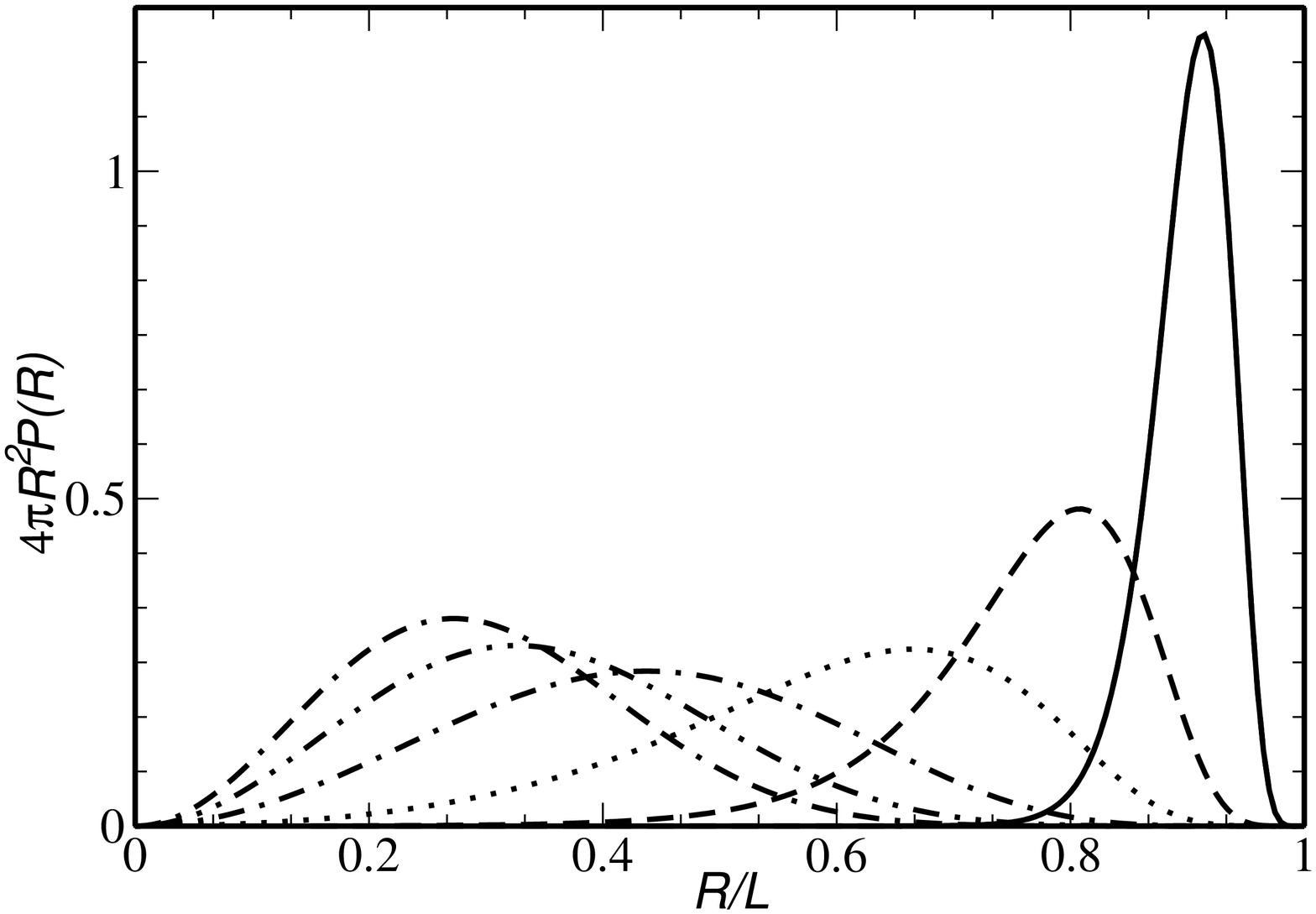}

\begin{flushleft}FIG. 3. Normalized polymer propagator $4\pi R^{2}P\left(R\right)$
versus $R/n$ for $n=10$. Continuous line $\left(\alpha=0.01\right)$
, dashed line $\left(\alpha=0.02\right)$, dotted line $\left(\alpha=0.03\right)$,
dashed-dotted line $\left(\alpha=0.05\right)$, dashed-dotted-dotted
line $\left(\alpha=0.1\right)$ and dashed-dashed-dotted line $\left(\alpha=0.75\right)$.\end{flushleft}
\end{figure}

\clearpage

Comment: Figure 4, First Author: Marcelo Marucho, JCP

\begin{figure}
\includegraphics[%
  width=4in,
  keepaspectratio,
  angle=270]{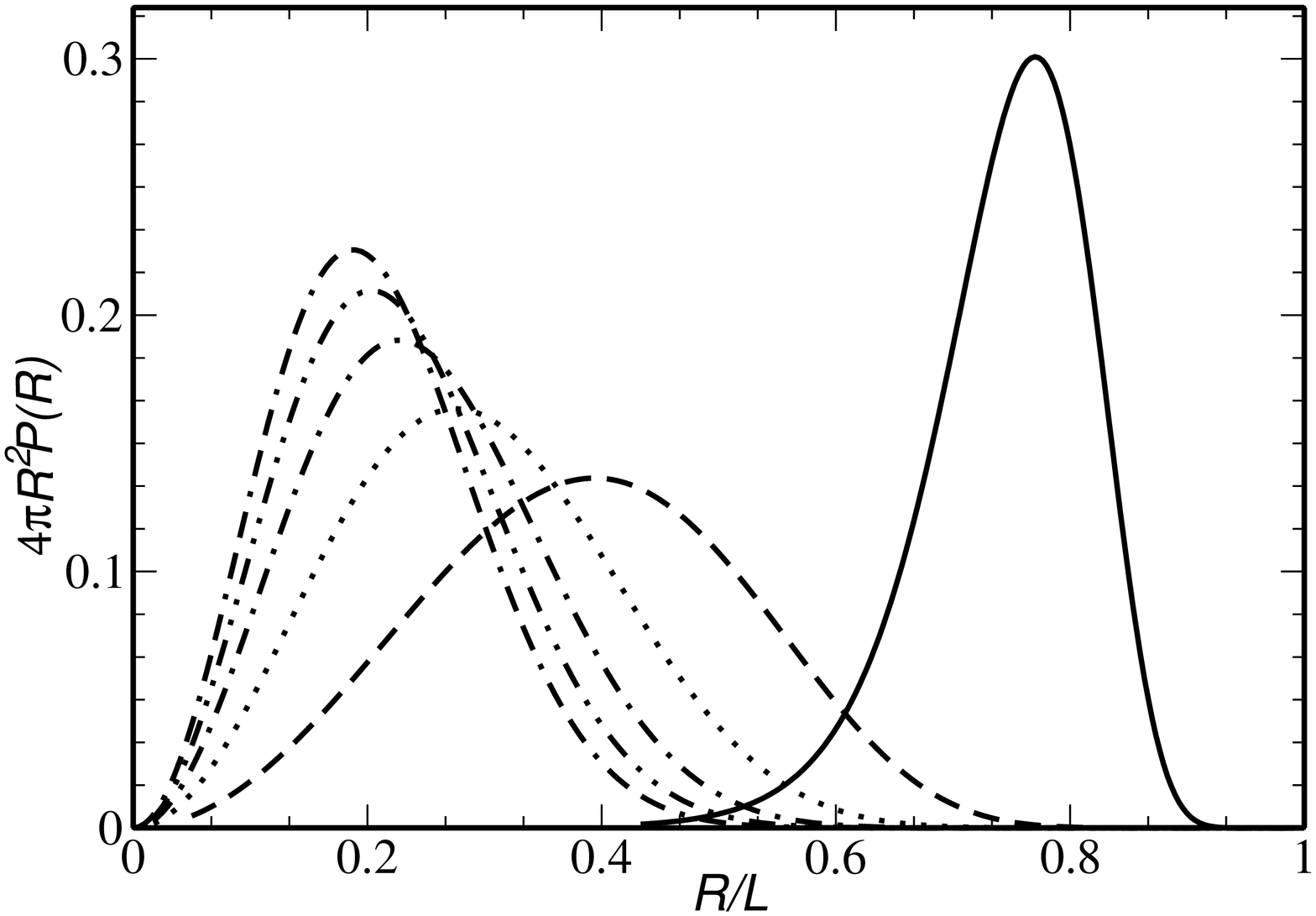}

\begin{flushleft}FIG. 4. Normalized polymer propagator $4\pi R^{2}P\left(R\right)$
versus $R/n$ for $n=20$. Continuous line $\left(\alpha=0.01\right)$
, dashed line $\left(\alpha=0.02\right)$, dotted line $\left(\alpha=0.03\right)$,
dashed-dotted line $\left(\alpha=0.05\right)$, dashed-dotted-dotted
line $\left(\alpha=0.1\right)$ and dashed-dashed-dotted line $\left(\alpha=0.75\right)$.\end{flushleft}
\end{figure}

\clearpage

Comment: Figure 5, First Author: Marcelo Marucho, JCP

\begin{figure}
\includegraphics[%
  width=4in,
  keepaspectratio,
  angle=270]{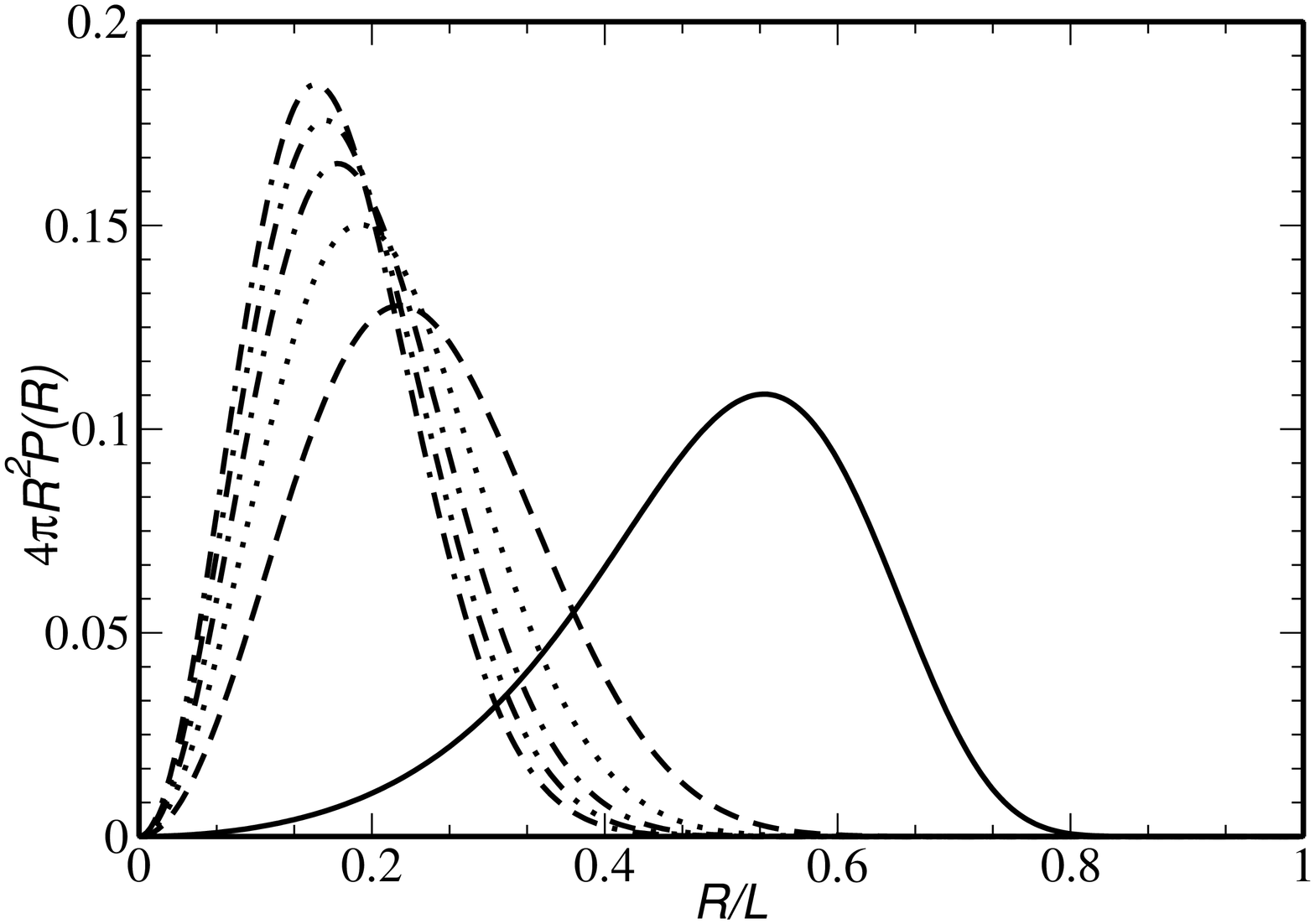}

\begin{flushleft}FIG. 5. Normalized polymer propagator $4\pi R^{2}P\left(R\right)$
versus $R/n$ for $n=30$. Continuous line $\left(\alpha=0.01\right)$
, dashed line $\left(\alpha=0.02\right)$, dotted line $\left(\alpha=0.03\right)$,
dashed-dotted line $\left(\alpha=0.05\right)$, dashed-dotted-dotted
line $\left(\alpha=0.1\right)$ and dashed-dashed-dotted line $\left(\alpha=0.75\right)$.\end{flushleft}
\end{figure}

\clearpage

Comment: Figure 6, First Author: Marcelo Marucho, JCP

\begin{figure}
\includegraphics[%
  width=4in,
  keepaspectratio,
  angle=270]{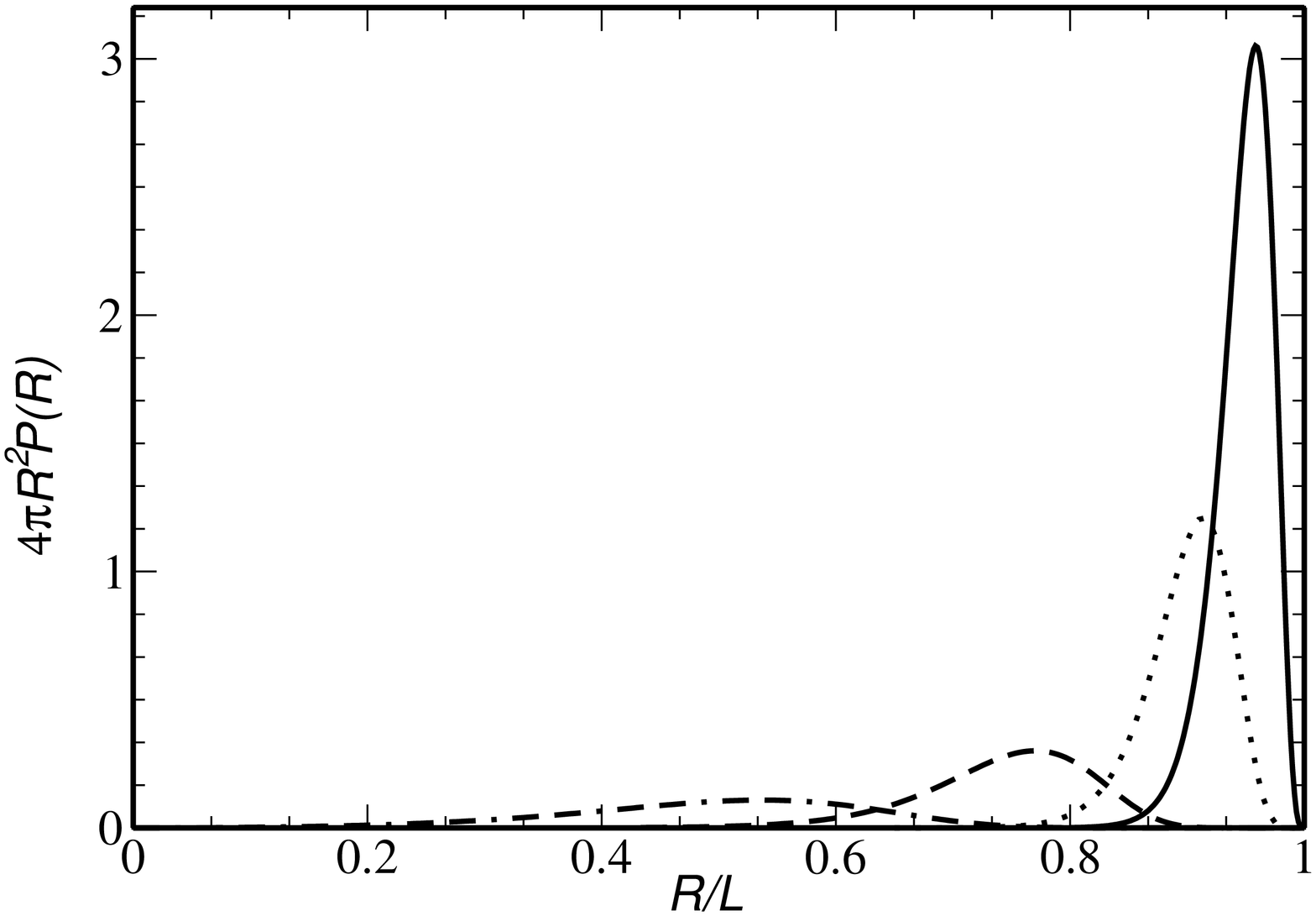}

\begin{flushleft}FIG. 6. Normalized polymer propagator $4\pi R^{2}P\left(R\right)$
versus $R/n$ for $\alpha=0.01$. Continuous line $\left(n=6\right)$,
dotted line $\left(n=10\right)$, dashed line $\left(n=20\right)$
and dashed-dashed-dotted line $\left(n=30\right)$.\end{flushleft}
\end{figure}

\clearpage

Comment: Figure 7, First Author: Marcelo Marucho, JCP

\begin{figure}
\includegraphics[%
  width=4in,
  angle=270]{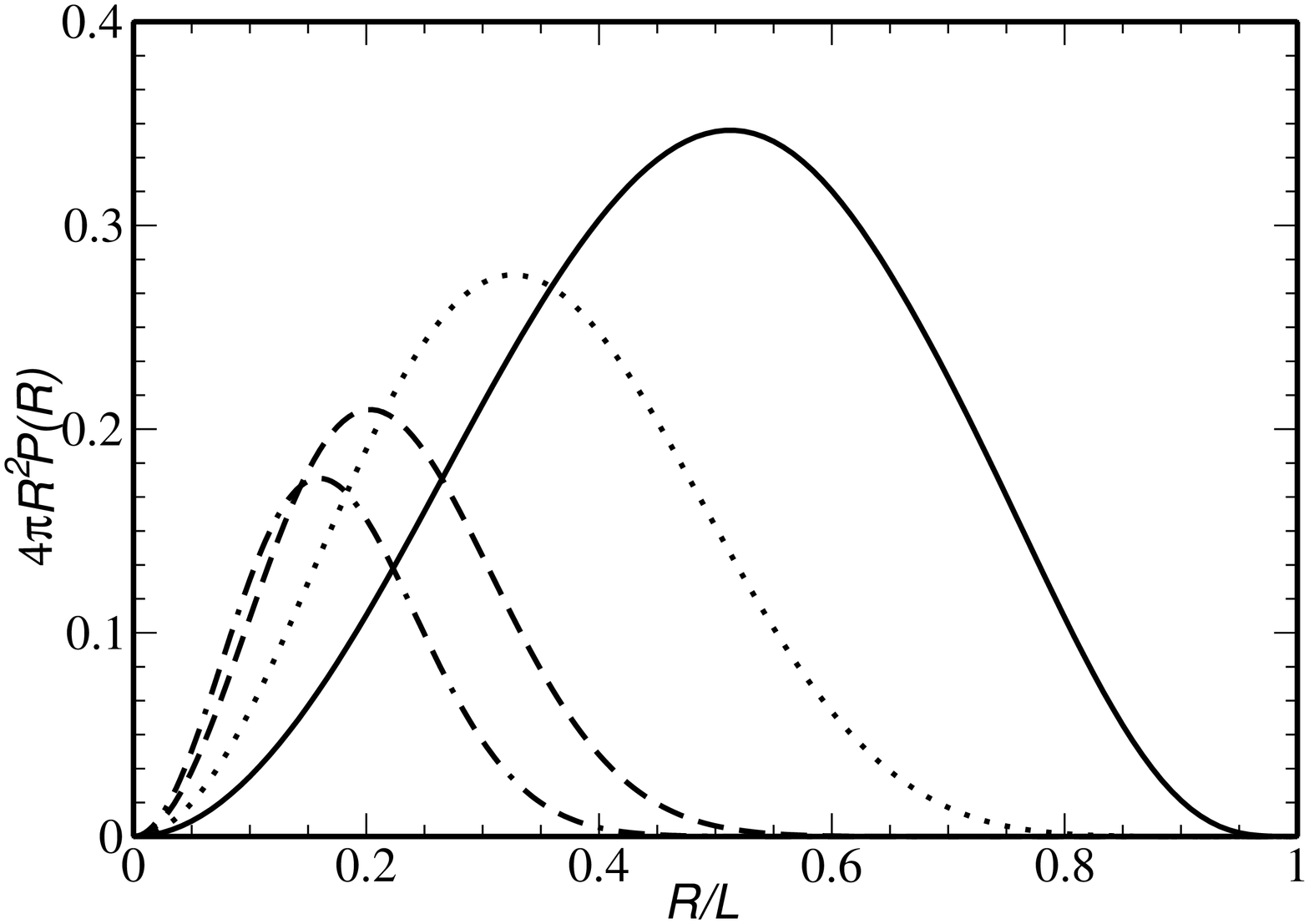}

\begin{flushleft}FIG. 7. Normalized polymer propagator $4\pi R^{2}P\left(R\right)$
versus $R/n$ for $\alpha=0.1$. Continuous line $\left(n=6\right)$,
dotted line $\left(n=10\right)$, dashed line $\left(n=20\right)$
and dashed-dashed-dotted line $\left(n=30\right)$.\end{flushleft}
\end{figure}

\clearpage

Comment: Figure 8, First Author: Marcelo Marucho, JCP

\begin{figure}
\includegraphics[%
  width=4in,
  angle=270]{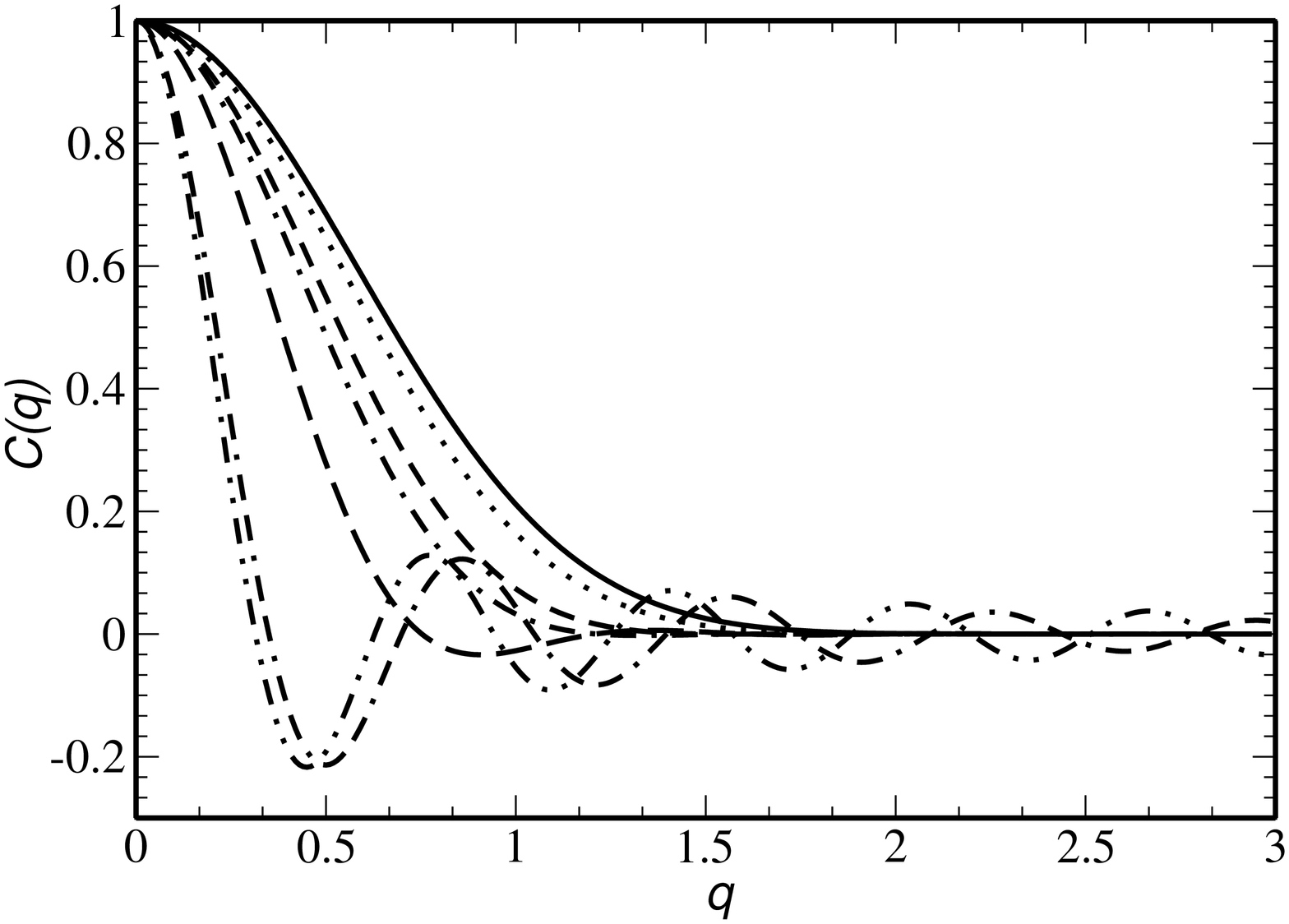}

\begin{flushleft}FIG. 8. Characteristic function $C\left(q\right)$
versus wave vector $q$ for $n=10$. Dashed-dotted-dotted line $\left(\alpha=0\right)$
(the exact solution of rigid Model), dashed-dashed-dotted line $\left(\alpha=0.01\right)$,
long dashed line $\left(\alpha=0.04\right)$, dashed-dotted line $\left(\alpha=0.07\right)$,
dashed line $\left(\alpha=0.1\right)$, dotted line $\left(\alpha=0.75\right)$
and continuous line $\left(\alpha=\infty\right)$ (the exact solution
of the Random Flight Model).\end{flushleft}
\end{figure}

\clearpage

Comment: Figure 9, First Author: Marcelo Marucho, JCP

\begin{figure}
\includegraphics[%
  width=4in,
  keepaspectratio,
  angle=270]{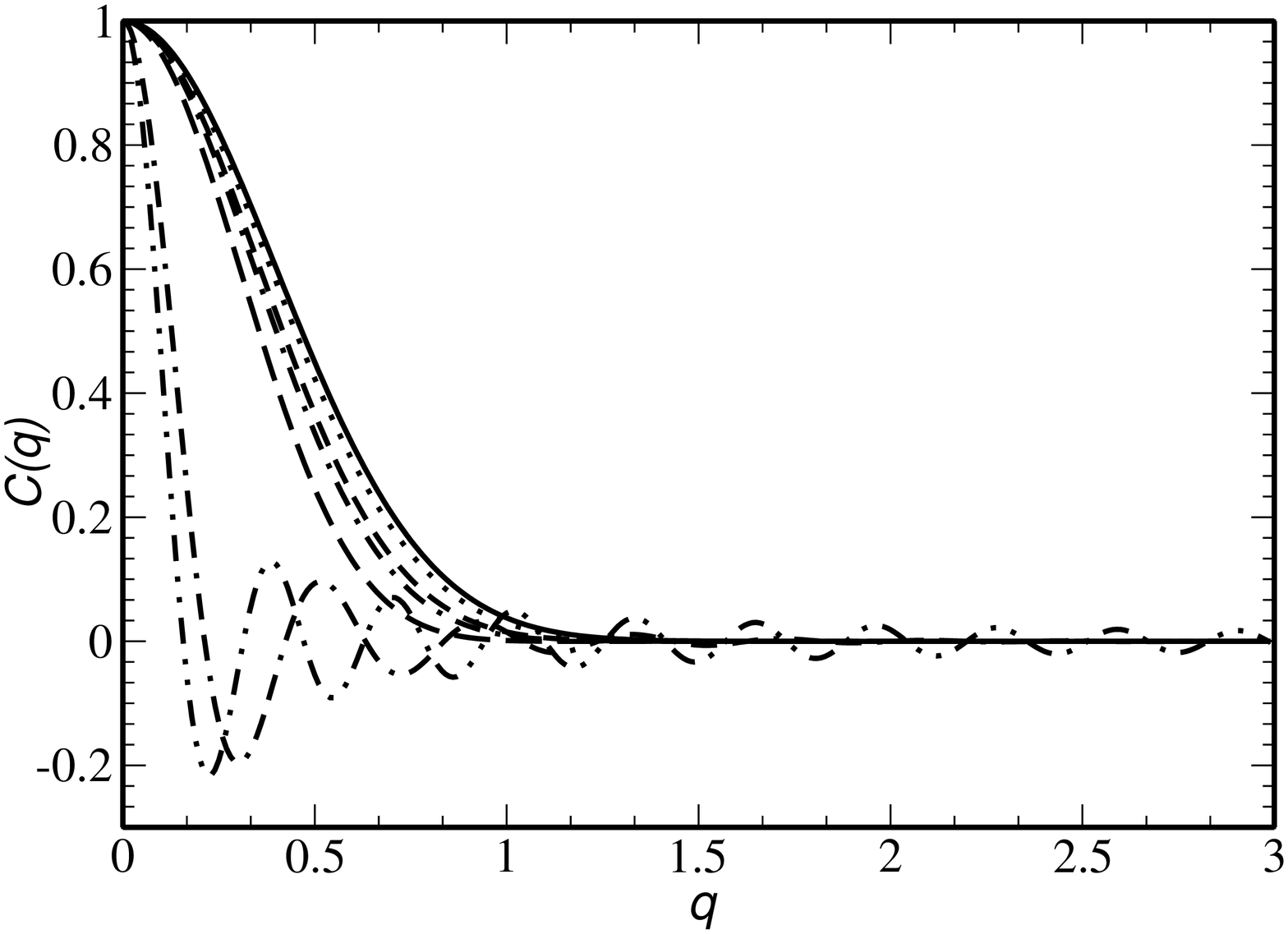}

\begin{flushleft}FIG. 9. Characteristic function $C\left(q\right)$
versus wave vector $q$ for $n=20$. Dashed-dotted-dotted line $\left(\alpha=0\right)$
(the exact solution of rigid Model), dashed-dashed-dotted line $\left(\alpha=0.01\right)$,
long dashed line $\left(\alpha=0.04\right)$, dashed-dotted line $\left(\alpha=0.07\right)$,
dashed line $\left(\alpha=0.1\right)$, dotted line $\left(\alpha=0.75\right)$
and continuous line $\left(\alpha=\infty\right)$ (the exact solution
of the Random Flight Model).\end{flushleft}
\end{figure}

\clearpage

Comment: Figure 10, First Author: Marcelo Marucho, JCP

\begin{figure}
\includegraphics[%
  width=4in,
  keepaspectratio,
  angle=270]{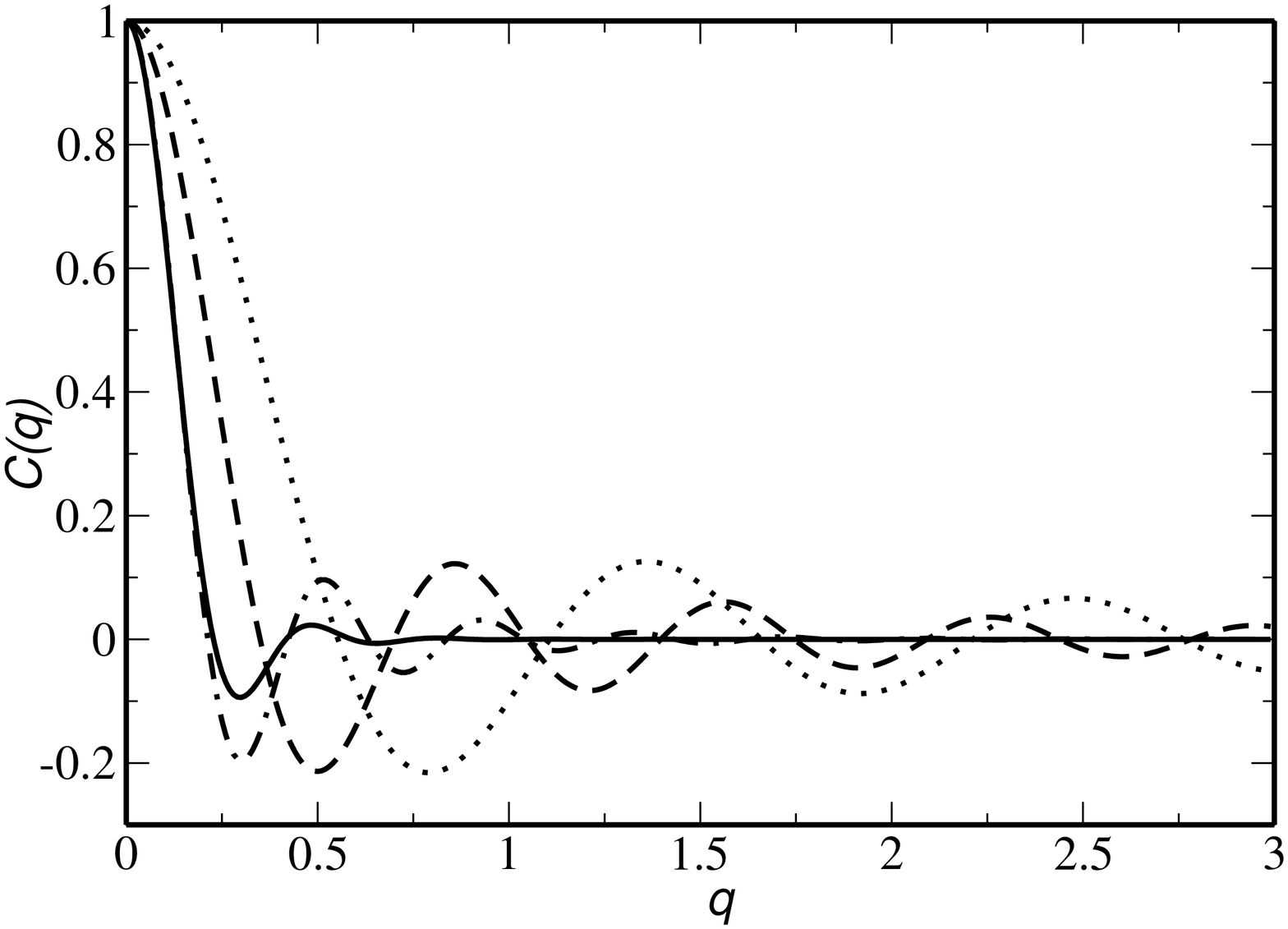}

\begin{flushleft}FIG. 10. Characteristic function $C\left(q\right)$
versus wave vector $q$ for $\alpha=0.01$. Dotted line $\left(n=6\right)$,
dashed line $\left(n=10\right)$, dashed-dotted line $\left(n=20\right)$
and continuous line $\left(n=30\right)$.\end{flushleft}
\end{figure}

\clearpage

Comment: Figure 11, First Author: Marcelo Marucho, JCP

\begin{figure}
\includegraphics[%
  width=4in,
  keepaspectratio,
  angle=270]{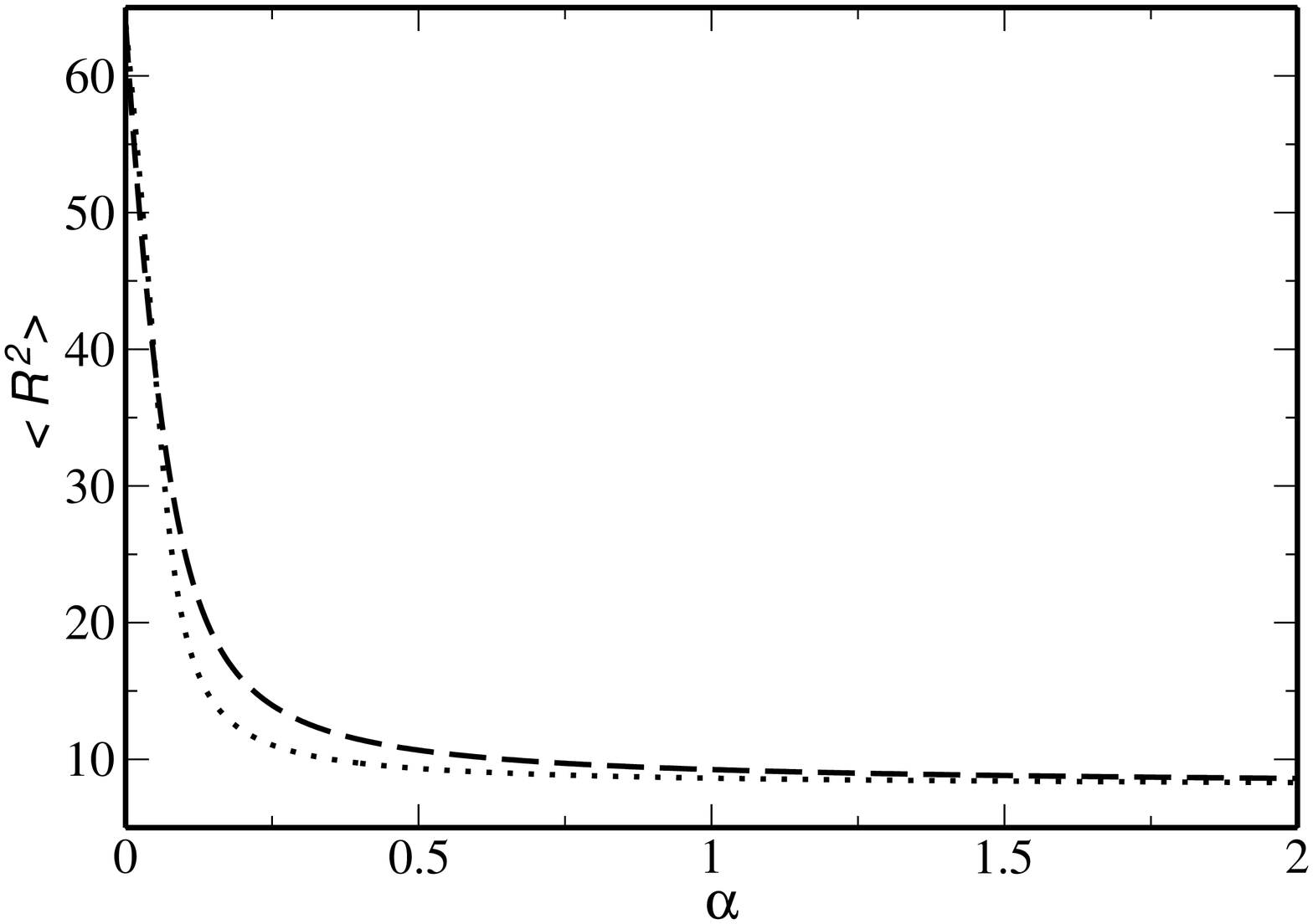}

\begin{flushleft}FIG. 11. Mean squared end-to-end distance $\left\langle \mathbf{R}^{2}\right\rangle $
versus the parameter $\alpha$ for $n=8$. The dotted line is our
approximate solution and the dashed line is the exact solution of
the KP model\cite{Benoit}.\end{flushleft}
\end{figure}

\clearpage

Comment: Figure 12, First Author: Marcelo Marucho, JCP

\begin{figure}
\includegraphics[%
  width=4in,
  keepaspectratio,
  angle=270]{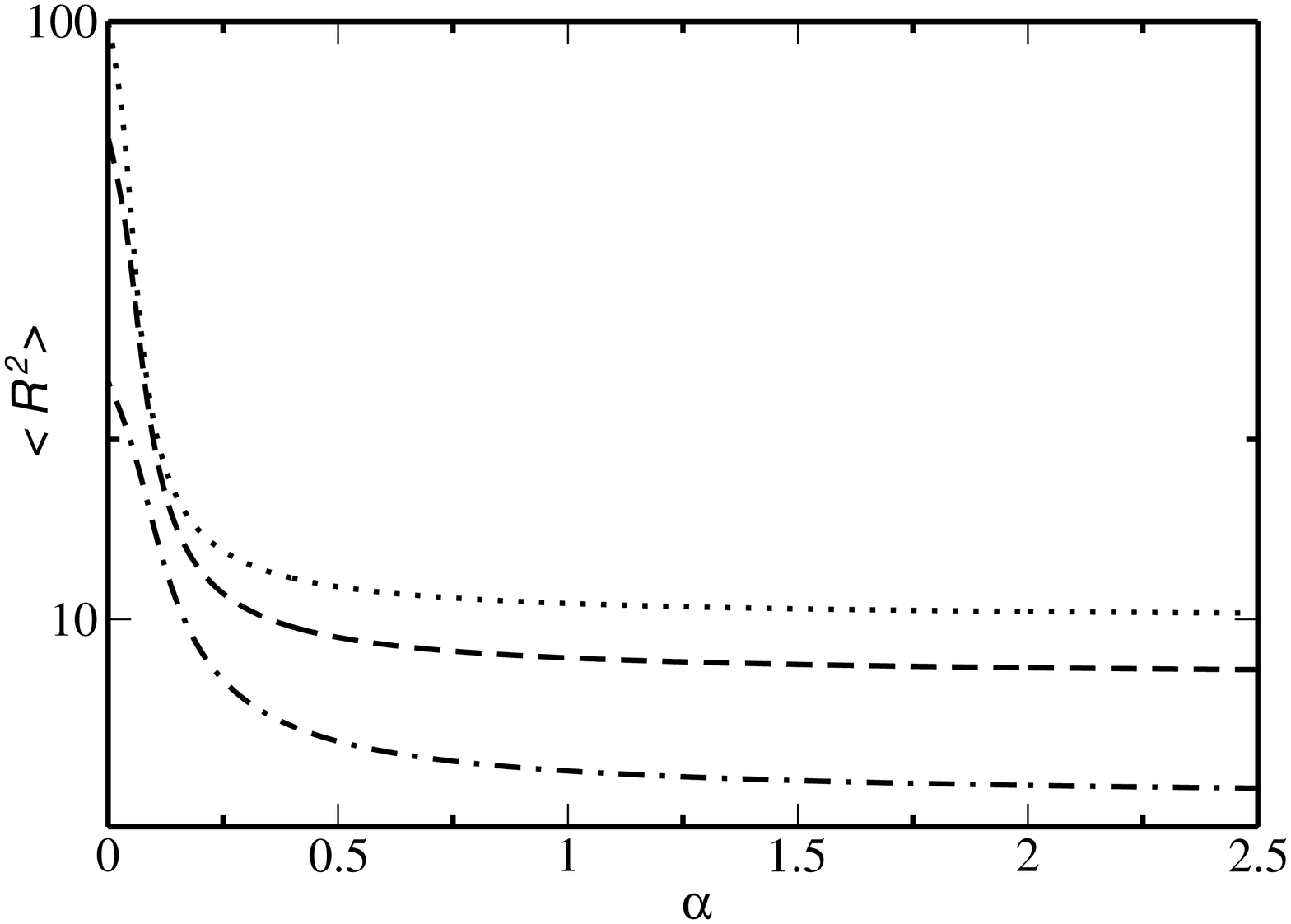}

\begin{flushleft}FIG. 12. Mean square end-to-end distance $\left\langle \mathbf{R}^{2}\right\rangle $
(in logarithmic scale) versus the parameter $\alpha$. Dashed-dotted
line $\left(n=5\right)$, dashed line $\left(n=8\right)$ and dotted
line $\left(n=10\right)$.\end{flushleft}
\end{figure}

\clearpage

Comment: Figure 13, First Author: Marcelo Marucho, JCP

\begin{figure}
\includegraphics[%
  width=4in,
  keepaspectratio,
  angle=270]{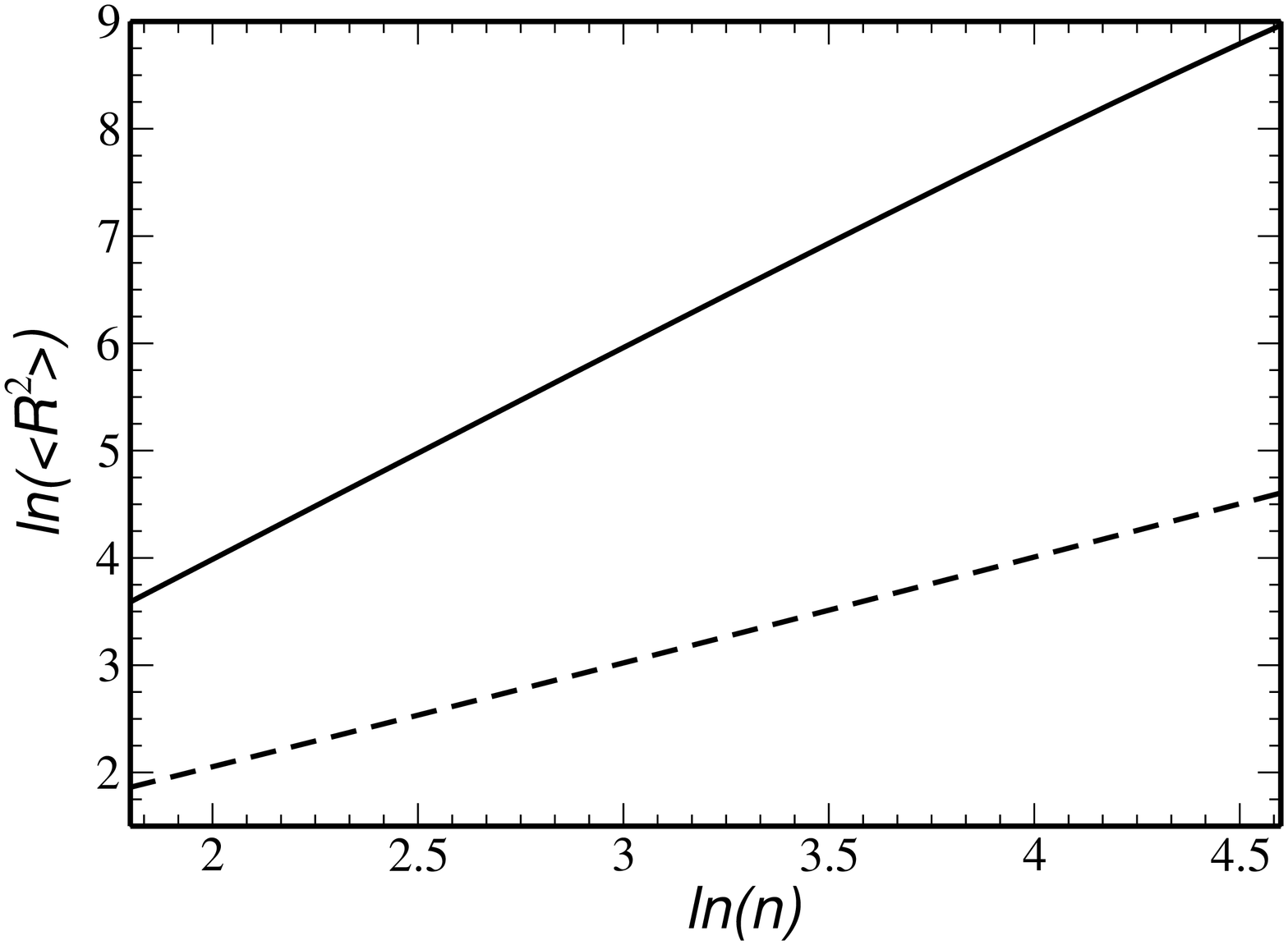}

\begin{flushleft}FIG. 13. $ln\left\{ \left\langle \mathbf{R}^{2}\right\rangle \right\} $
versus the number of segments $ln\left(n\right)$. Continuous line
$(\alpha=0.001)$ and dashed line $(\alpha=0.75)$.\end{flushleft}
\end{figure}

\clearpage
\end{document}